\documentclass{aa}

\usepackage{graphicx}
\usepackage[dvips]{color}
\usepackage{txfonts}
\newcommand{\abbrev}[1]{{#1}}      %%  plain, final
\newcommand{\distsag}{{25.14}}
\newcommand{\ebv}{$E(B-V)$}
\newcommand{\dmod}{$(m-M)_{\circ}$}
\newcommand{\ebvsag}{{0.12}}
\newcommand{\fehsag}{{-2.21}}

\newcommand{\vv}{{$m_{\rm F606W}$}}
\newcommand{\ii}{{$m_{\rm F814W}$}}
\newcommand{\bbvv}{{$m_{\rm F475W} - m_{\rm F606W}$}}
\newcommand{\bbii}{{$m_{\rm F475W} - m_{\rm F814W}$}}
\newcommand{\vvii}{{$m_{\rm F606W} - m_{\rm F814W}$}}
\newcommand{\vviiZW}{{$m_{\rm F555W} - m_{\rm F814W}$}}
\newcommand{\allstar}{{\sc allstar}}

\newcommand{\besancon}{Besan\c{c}on}
\newcommand{\hst}{{\sl HST}}
\newcommand{\zwi}{I\,Zw\,18}

\newcommand{\hi}{\ion{H}{i}}
\newcommand{\hii}{\ion{H}{ii}}

\newcommand{\ism}{{ISM}}

\newcommand{\kms}{km s$^{\rm -1}$}

\begin{document}

\title{HST/ACS observations of the old and  metal-poor \\ Sagittarius dwarf 
irregular galaxy 
\thanks{Based on observations with the NASA/ESA {\it Hubble Space
Telescope}, (GO 9820) obtained at the Space Telescope Science
Institute, which is operated by AURA, Inc., under NASA contract NAS
5-26555.}  }

\author{Y. Momany\inst{1}
        \and 
        E. V. Held\inst{2} 
        \and 
        I. Saviane\inst{3} 
        \and 
        L. R. Bedin\inst{1,4}
        \and 
        M. Gullieuszik\inst{1,2} 
        \and 
        M. Clemens\inst{2} 
        \and \\
	L. Rizzi\inst{5}
        \and 
	M. R. Rich\inst{6}
        \and 
	K. Kuijken\inst{7} 
	}
\offprints{Y. Momany, \email{momany@pd.astro.it}}

\institute {Dipartimento di Astronomia, Universit\`a di Padova, Vicolo
dell'Osservatorio 2, I-35122 Padova, Italy
\and Osservatorio Astronomico di Padova, Vicolo dell'Osservatorio 5,
I-35122 Padova, Italy
\and European Southern Observatory, 3107 Alonso de Cordova, Vitacura,
Casilla 19001, Santiago 19, Chile
\and European Southern Observatory, Karl-Schwarzschild-Str. 2, 85748
Garching b. M\"unchen, Germany
\and 
Institute for Astronomy, 2680 Woodlawn Drive, Honolulu, USA
\and 
Department of Physics and Astronomy, UCLA, 430 Portola Plaza, Los
Angeles CA, 90095-1547
%
%Department of Astronomy, University of California Los Angeles,
%Los Angeles, CA 90095, USA
%
\and
Leiden Observatory, PO Box 9513, NL-2300 RA Leiden, The Netherlands
}

\date{Received \dots; accepted \dots}

\abstract{We have   obtained deep color-magnitude      diagrams (CMDs)
of the Sagittarius dwarf irregular galaxy with the Advanced Camera for
Surveys (ACS)  on board  the  {\it Hubble Space Telescope}.    The new
diagrams show for the first time the star formation history of SagDIG.
The young main-sequence is   traced  down to  \vv~$\simeq27.5$,   well
separated from the older He  burning blue loop  stars.  The wide color
range spanned  by the blue  and  red super-giants  is consistent  with
Padua evolutionary models with $Z\sim0.0004$.
A well-defined red giant branch indicates that the galaxy is dominated
by a stellar population older than 1 Gyr, for  which we have estimated
a metallicity in the range [Fe/H]~$= -2.2$ to  $-1.9$ depending on the
adopted reddening.
The identification of  several C stars  on a luminous asymptotic giant
branch, as well as  the presence of a   conspicuous red clump with  an
elongated structure, are    indicative of an  extended  star formation
epoch between 1 and 10 Gyr ago.
Perhaps most importantly,  we  have identified a small   population of
stars on a genuinely  old red horizontal branch at \vv$=25.70\pm0.05$,
which sets the first  epoch of star formation in  SagDIG back to $\sim
9-10$ Gyr ago. {\em This shows that SagDIG has been able to maintain a
very low mean metallicity over a 10 Gyr life time,  and that all Local
Group dwarf galaxies share an ancient ($\ge10$ Gyr) initial episode of
star formation}.
Thus, at the moment, \zwi~ remains  the only exception to this general
trend,  favoring the so-called ``young galaxy''  scenario where a very
low metallicity galaxy is not necessarily  old. However, a re-analysis
of ACS observations of \zwi~ suggests that an older RGB population may
be present also in \zwi.

In  order to  address the interplay  between   stars and Inter-Stellar
Medium, we compare  the distribution of atomic  hydrogen with  that of
stellar populations of various ages.  The youngest  stars are found in
clumps located:   (i) near to,  but  not coincident  with, the highest
column densities of \hi, and  (ii) in tails reminiscent of propagating
star formation.  We find,  however, little correlation between stellar
populations older  than $\sim 1$  Gyr  and the \hi.   In particular we
fail to find  any generation of stars  that preferentially lie  within
the large \hi \ hole.   We therefore exclude  the possibility that the
ring-like \hi\   distribution  is the  result   of multiple  supernova
events.  Alternative scenarios have to be explored for the ring-like
\hi~ morphology of SagDIG,  which amongst dwarf irregular galaxies, is
not unusual.

\keywords{galaxies: individual (SagDIG) --- individual (\zwi) ---
	  galaxies: structure --- galaxies: dwarf --- galaxies: photometry --- 
	  galaxies: Local Group --- galaxies: fundamental parameters ---
	  galaxies: ISM --- Galaxy: stellar content ---stars:
formation	} 
}

\maketitle

%--------------------------------------------------------------------
\section{Introduction}

The Hubble Space Telescope has made it possible to resolve the stellar
populations    of dwarf    irregular   galaxies (\abbrev{dIrr})   into
individual stars   both  in the Local   Group  (e.g.   Dohm-Palmer  et
al. \cite{dohm97}, \cite{dohm98}, \cite{dohm02}; Tolstoy et al.
\cite{tolstoy98}; Dolphin \cite{dolphin00}; Dolphin et al.
\cite{dolphin01}, \cite{dolphin02}, \cite{dolphin03}; Skillman et al.
\cite{skill03})
and beyond (as  in  the case  of the   blue-compact dwarf \zwi\   at a
distance between $10-15$~Mpc; Aloisi et al. \cite{aloisi99}; Izotov \&
Thuan   \cite{izo04}).    In  all   cases,  color-magnitude   diagrams
(\abbrev{CMD}) can be   constructed  to estimate their distance    and
metallicity and  constrain   the    recent  star   formation   history
(\abbrev{SFH}) of  the systems. In  the nearest  galaxies, a wealth of
evolutionary  details can be   observed,  otherwise hampered  by  poor
resolution in ground-based studies.
%

%--------------Introducing young! metal poor galaxies
%
The quest for {\it   very metal-poor galaxies} among dwarf  irregulars
started soon after  the Searle and Sargent (\cite{searle72}) discovery
that  the interstellar medium (\abbrev{ISM})  of I~Zw~18 has an oxygen
abundance that is $\sim1/50$ of the solar one.
As  Searle   \& Sargent  pointed  out,   this is  compatible with  two
scenarios:  either   the   galaxy  is  now  experiencing   its   first
star-formation (\abbrev{SF}) episode,  or, if it contains $\sim10$ Gyr
old stars,  its SF must  have proceeded in  strong bursts separated by
quiescent phases.  The first   hypothesis  (a {\em young} galaxy)   is
relevant  in  the context    of    cosmological models that     assume
hierarchical formation of structures  in a cold dark matter  dominated
Universe.   In that scenario, dwarf  dark matter  halos ($\sim 10^{8}$
$\rm M_{\odot }$)  would be the first objects  to condense out  of the
Hubble flow, and it is  therefore important to understand whether they
already contained stars before merging into larger galaxies.
In   fact,  \zwi\ is  a  difficult  object  to  study because  of  its
distance. We  may ask  ourselves   an equivalent, possibly  easier  to
answer,  question: {\it are there any  galaxies with a very metal-poor
young population and, at the same  time, a $\sim10$ Gyr old generation
of   stars} ?  The existence  of  such old, metal-poor stellar systems
would imply  that very low metal abundances  are not inconsistent with
an ancient epoch of galaxy formation.
%
%-----------Introducing SagDIG: Literature
%
In order to answer this question, we have undertaken an HST/ACS study
the Sagittarius dwarf irregular galaxy (also known as SagDIG or
UKS\,1927--177). The interest of SagDIG arises from the presence of
on-going star formation in an extremely metal-poor environment.  The
fact that SagDIG is a {\it nearby} ($\simeq1.1$ Mpc) galaxy makes it
an ideal target to test locally the {young galaxy} hypothesis.

%-------- PHOTOMETRY -----------
The  most  recent photometric studies     of   SagDIG are  those    of
Karachentsev et al.   (\cite{kara99}), Lee \& Kim (\cite{lee00}),  and
Momany et al.  (\cite{momany02}, hereafter M02).
Lee \& Kim  and Karachentsev and  coll. concluded that  SagDIG has the
lowest   stellar metallicity (up  to   [Fe/H]$\sim -2.45$) among Local
Group star-forming galaxies. M02   presented deep $BVI$  photometry of
SagDIG obtained  from ESO NTT data (we  refer the reader to that paper
for a full summary of previous work).
In M02 a differential reddening scenario was suggested to explain the
relatively red color of main-sequence  stars. Our hypothesis was  that
some  internal reddening  (in  addition to  the foreground  reddening)
affects  the    young stars in   star-forming    regions.  The revised
metallicity,  [Fe/H]$=  -2.1 \pm  0.2$, turned   out to  be  in better
agreement     with   the  general       trend    of  the     (stellar)
metallicity--luminosity  relation   for dwarf   irregular    galaxies,
indicating a ``normal'', although metal-poor, dIrr galaxy.

% --- INSTERSTELLAR MEDIUM ------
The cold and warm \ism\ of SagDIG and its kinematics have been the
subject of several investigations.
Skillman et al. (\cite{skill89}) obtained optical spectrophotometry of
the  most luminous  \hii\ regions,   and  estimated a [O/H]  abundance
$\sim3\%$ of the solar value, a result  independently confirmed by the
new measurements of Saviane et al. (\cite{saviane02}).
%
%--- (O/H) for         SagDIG is = 2.6 \pm 1.3 \times 10^{-5}
%--- (O/$H_{\odot}$) Solar is = 8.3 \times         10^{-4}  
%
%

High-resolution, high sensitivity VLA observations  of the SagDIG \hi\
content  have been obtained by Young  \& Lo (\cite{young97}) (see also
Lo  et  al.   \cite{lo93}).   About  $1.2 \times 10^7$  M$_{\odot}$ of
(H\,$+$\,He)  have been estimated by   M02 using \dmod~$=\distsag$ and
the \hi\ column density given by Young \& Lo.
%-- 
SagDIG thus  appears to have a  relatively  high mass fraction  in the
form of neutral  gas, $M_{H I}/L_B= 1.6$, a  value typical of the \hi\
content in dwarf irregular galaxies (see Mateo \cite{mateo98}).
The   \hi\   is distributed in   an  almost  symmetric  ring which was
interpreted  as produced by the  combined effects of stellar winds and
supernovae.
%

%-   oooooooo WHY HST ??? oooooooooo
%------------ THE PAPER ------------
This paper presents our  new HST/ACS observations,  clearly overcoming
the limitations of previous  ground-based observations and probing the
young, intermediate age, and old stellar populations in SagDIG.
The  results of those  observations are  presented in  this paper. The
paper   is organized as   follows.  In  Sect.~\ref{S_observations}  we
introduce    our  HST/ACS    observations     and    data   reduction.
Section~\ref{S_CMD}   presents the  ACS  color-magnitude  diagrams and
their main features.
In Sect.~\ref{S_feh} we re-derive the distance to SagDIG, the mean
metallicity of its red giants stars, and, most importantly, provide 
evidence for a 10 Gyr old horizontal branch population.
The recent star formation is then analyzed in Sect.~\ref{S_recentSF}.
Section~\ref{S_HI} compares the spatial distribution of the stars and
the \hi\ and discusses the relationship between the composite stellar
populations of SagDIG and the morphology of its ISM.
The principal results of our ACS study are finally summarized in
Sect.~\ref{S_discuss}.

%-------------------------------------------------------------
\begin{figure}[hb!]
\centering \resizebox{\hsize}{!}{\includegraphics{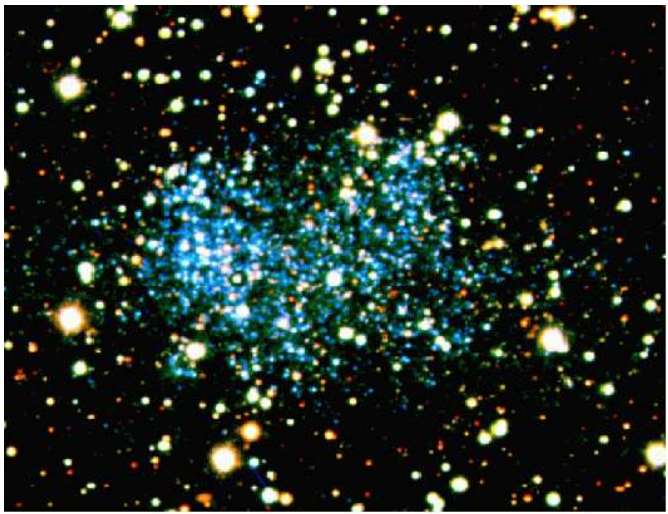}}
%\centering \resizebox{\hsize}{!}{\includegraphics{./psfiles/F_SagDIG_NTT.ps}}
%\centering \includegraphics[width=7cm,height=7cm]{./psfiles/F_SagDIG_NTT.ps}
\caption{A $3.1\times2.4$ arcmin true-color image of SagDIG derived
from  our  ground-based    ESO/NTT $BVI$   observations     (1\arcsec\
resolution).  North is   to the  top, East  to    the left.   The  low
resolution of NTT data highlights large-scale  structures such as {\it
tails} of massive hot stars and regions almost devoid of young stars.}
\label{f_ntt}
\end{figure}
%-------------------------------------------------------------

%%%%%%%%%%%%%%%%%%%%%%%%%%%%%%%%%%%%%%%%%%%%%%%%%%%%%%%%%%%%%%%%%%%%%%
%%								    %%
\section{Observations and data reduction}			    %%
\label{S_observations}
%%								    %%
%%%%%%%%%%%%%%%%%%%%%%%%%%%%%%%%%%%%%%%%%%%%%%%%%%%%%%%%%%%%%%%%%%%%%%

%--------------------------------------------------------------->
\begin{table}[bh!]
\caption{ Photometric errors and completeness as derived from
artificial star experiments. \label{t_complet}}
\centering
\begin{tabular}{lcccccc}
\hline\hline 
 mag & $\sigma_{F475W}$ & $C_{F475W}$ & $\sigma_{F606W}$ & $C_{F606W}$ & 
$\sigma_{F814W}$ & $C_{F814W}$ \\
\hline
 20.00 &  0.010 &  1.000 &  0.010 &  1.000 &  0.010 &  1.000  \\
 20.40 &  0.010 &  1.000 &  0.010 &  1.000 &  0.010 &  1.000  \\
 20.80 &  0.010 &  1.000 &  0.010 &  1.000 &  0.009 &  0.999  \\
 21.20 &  0.010 &  1.000 &  0.010 &  1.000 &  0.009 &  0.999  \\
 21.60 &  0.010 &  1.000 &  0.010 &  1.000 &  0.011 &  0.999  \\
 22.00 &  0.010 &  1.000 &  0.010 &  1.000 &  0.013 &  0.997  \\
 22.40 &  0.010 &  1.000 &  0.008 &  1.000 &  0.013 &  0.996  \\
 22.80 &  0.010 &  1.000 &  0.012 &  0.999 &  0.016 &  0.996  \\
 23.20 &  0.009 &  1.000 &  0.014 &  1.000 &  0.021 &  0.994  \\
 23.60 &  0.012 &  1.000 &  0.016 &  0.999 &  0.025 &  0.991  \\
 24.00 &  0.019 &  1.000 &  0.019 &  0.997 &  0.033 &  0.988  \\
 24.40 &  0.027 &  0.996 &  0.023 &  0.996 &  0.043 &  0.989  \\
 24.80 &  0.036 &  0.993 &  0.028 &  0.994 &  0.052 &  0.990  \\
 25.20 &  0.046 &  0.989 &  0.036 &  0.992 &  0.066 &  0.963  \\
 25.60 &  0.060 &  0.988 &  0.047 &  0.991 &  0.086 &  0.894  \\
 26.00 &  0.077 &  0.974 &  0.059 &  0.988 &  0.114 &  0.801  \\
 26.40 &  0.097 &  0.932 &  0.076 &  0.978 &  0.148 &  0.715  \\
 26.80 &  0.125 &  0.807 &  0.102 &  0.966 &  0.186 &  0.603  \\
 27.20 &  0.167 &  0.677 &  0.137 &  0.924 &  0.225 &  0.384  \\
 27.60 &  0.211 &  0.428 &  0.175 &  0.552 &  0.263 &  0.168  \\
 28.00 &  0.256 &  0.182 &  0.212 &  0.141 &  0.302 &  0.047  \\
 28.40 &  0.301 &  0.043 &  0.249 &  0.008 &  0.341 &  0.008  \\
 28.80 &  0.345 &  0.010 &  0.286 &  0.009 &  0.379 &  0.004  \\
\hline
\end{tabular}
\end{table}
%---------------------------------------------------------------|

%-------------------------------------------------------------
\begin{figure*}[ht!]
\centering \includegraphics[width=18cm]{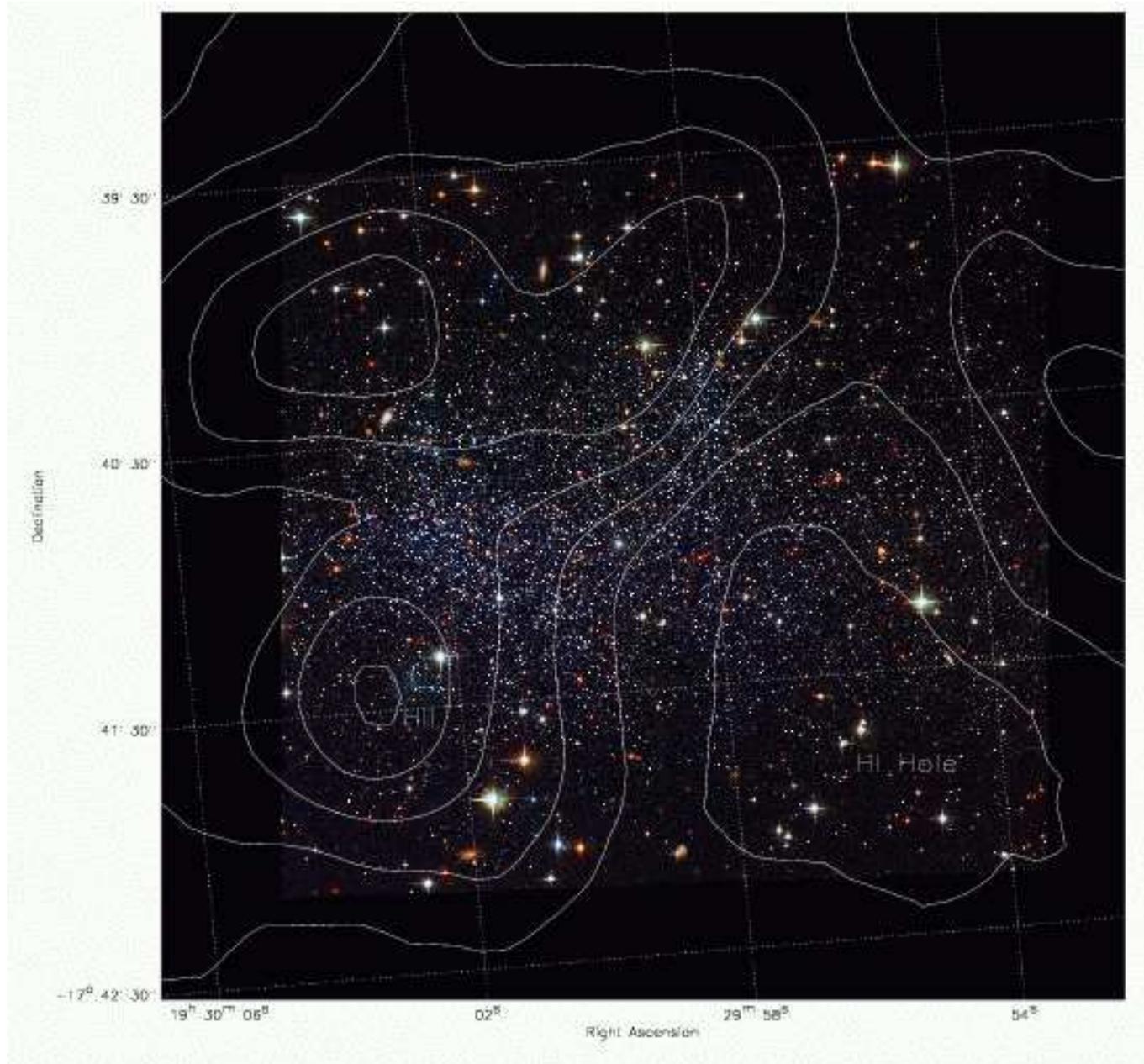}
%\centering \includegraphics[width=18cm]{./psfiles/HI_col.ps}
%\centering \includegraphics[width=18cm]{./psfiles/HI_CMYK_marco.ps}
%
\caption{True-color image of SagDIG derived from our ACS observations 
(see  also   {\sf  http://heritage.stsci.edu/2004/31/index.html}).   A
central, asymmetrical region   of  recent star formation,  with  faint
extensions, is superimposed on a sheet of red giant stars belonging to
the older stellar population of SagDIG.
Superimposed are  the \hi\  column  density contours  from Young \& Lo
(1997). Also plotted are the locations of the
\hi\  hole   and  \hii\  region.}
\label{f_acs}
\end{figure*}
%-------------------------------------------------------------

Images of SagDIG were taken  using the ACS on  board the \hst, in  the
course of one visit (on August 18, 2003; program  GO 9820) for a total
of 3  orbits.  The observed  field covers the main  body of the galaxy
and is centered at $l=21\fdg06$,  $b=-16\fdg28$.  Through the ACS-Wide
Field Channel (WFC), we  obtained  5$\times$396~s images in the  F606W
filter,  5$\times$419~s  in F475W  and  5$\times$421~s in  F814W.  The
series  of 5 images  in F475W and  F606W were dithered  so as to allow
filling  the gap  between  the  CCDs,  removal   of cosmic  rays,  and
optimization of the point  spread function (PSF) and  spatial sampling
(Anderson \& King
\cite{jay00}).
No attention has been paid in  the recovering the  gap in F814W images
(sub-arcsec dithering    pattern), so as  to   avoid  residuals in the
geometric distortion correction when second  epoch F814W images (to be
obtained in cycle 14) are used to derive the proper motion of observed
stars.

The instrument    distortion was corrected   using ACS/WFC geometrical
distortion correction maps made by  Anderson (2004, priv. comm.).  All
images in the  same filter were then  stacked to build a median image,
with pixel oversampling  by a factor of  2.  These steps allowed us to
improve the signal-to-noise ratio and the PSF sampling.
%.
%
%
Using these images, standard DAOPHOT/ALLSTAR (Stetson
\cite{stetson87}) routines were employed to derive both the PSF
magnitudes and their aperture corrections.
A  quadratically  variable PSF was   adopted  to account for  the  PSF
variations across the field of ACS (Krist \cite{kris03}).
The photometric calibration of  the ACS/WFC  filters follows Bedin  et
al.   (\cite{bedin04}).   This  zero-point  photometric calibration is
based on the Vega spectrum and  follows the recipes given by Holtzmann
et al. (\cite{holt+95})  for WFPC2, using  the  most updated in-flight
characterization of the camera (Sirianni et al. \cite{siri+04}).

Photometric  errors and  completeness  were estimated   from extensive
artificial stars experiments  following procedures similar to those of
M02.  More than  50,000 stars were  added to the  stacked images  in a
range  of magnitudes and   colors matching the  distribution of SagDIG
stars  in the observed  color-magnitude diagrams. The artificial stars
were placed  in a  suitable pattern  of   triangular meshes  to  avoid
self-crowding (see Saviane et al.  \cite{saviane00}).  The images with
the artificial  stars added were processed  exactly in the same way as
the original images   (in particular photometry  and calibration  were
done using the same parameters) and the photometry catalog was matched
with the input artificial star list.   {It should be noticed that this
procedure  is  approximated since   it  does not  fully reproduce  the
spatial variations  of   the PSF. Still, it   should   account for the
typical  errors  and completeness  of   the  data.}  The results   for
photometric    errors and completeness are   presented  in the form of
completeness contours in the color-magnitude  diagrams (see below) and
in Table~\ref{t_complet}.

Figure~\ref{f_ntt}  shows   a color-image  of  our ground-based  study
(M02), highlighting some large-scale patterns like  tails and holes in
the distribution of the young  blue stars.  Figure~\ref{f_acs} shows a
true color image of SagDIG derived from our  HST/ACS data. The spatial
resolution  of ACS  allows   us to resolve   both  the  young  and old
component in SagDIG. It also confirms the large-scale patterns seen in
ground-based images.

%-----------------------------------------------
\begin{figure*}
%\centering \includegraphics[width=1.0\textwidth,height=21cm]{./psfiles/F_CMDs.ps}
%\centering \includegraphics[width=0.9\textwidth]{./psfiles/F_CMDs.ps}
\centering \includegraphics[width=0.9\textwidth]{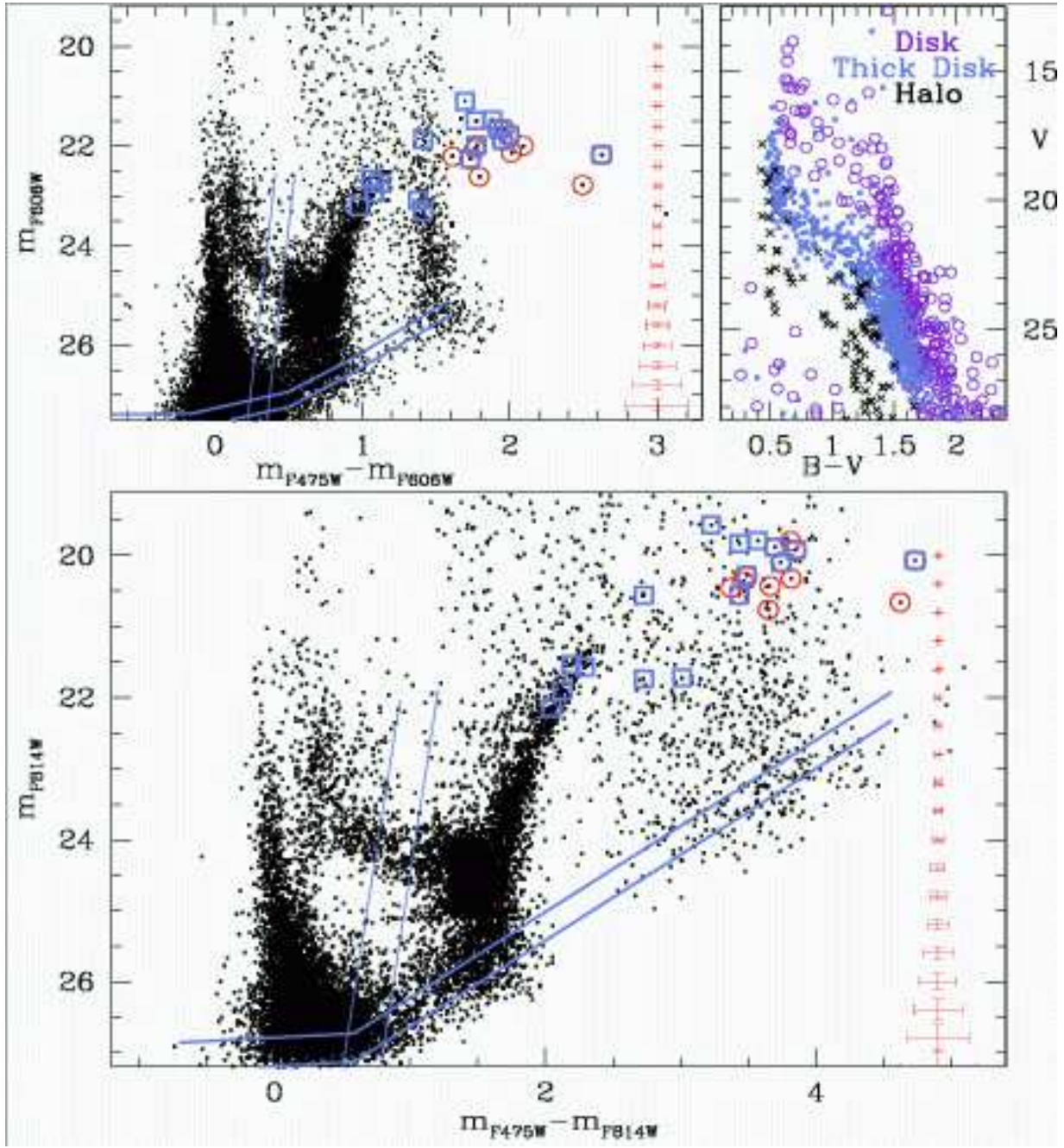}

\caption{HST/ACS color-magnitude diagrams of SagDIG. The {\it
upper-left}  panel shows the   \vv, (\bbvv)   diagram for  stars  with
$|SHARP|<0.3$.
The {\it thick lines} represent the 80\%  and 50\% completeness levels
as derived from artificial stars experiments. Mean error bars from the
same experiments   are plotted as  a  function of \vv.   The {\it open
symbols} represent the carbon stars identified by Demers
\& Battinelli (\cite{demers02}; {\it open circles}) and Cook
(\cite{cook87}; {\it open squares}).
The  {\it  thin lines} delimit     the  approximate location  of   the
instability strip   on the CMD.  The  {\it  upper-right} panel shows a
synthetic $V$, ($B-V)$\ diagram in the direction of SagDIG, calculated
using the WEB  interface of  the \besancon\  Galactic model (Robin  et
al. \cite{robin03}).  Different symbols refer  to the contributions of
the Galactic {thin disk ({\it open circles})}, {thick disk ({\it small
squares})} and {halo ({\it crosses})}. The {\it lower} panel shows the
\ii, (\bbii) diagram.}
\label{f_maincmd}
\end{figure*}
%-------------------------------------------------------------

%------------------------------------------one column figure
\begin{figure*}
\centering 
%\resizebox{\hsize}{!}{\includegraphics{./psfiles/F_Stellar_gradient.ps}}
%\includegraphics[width=13cm]{./psfiles/F_Stellar_gradient.ps}
\includegraphics[width=13cm]{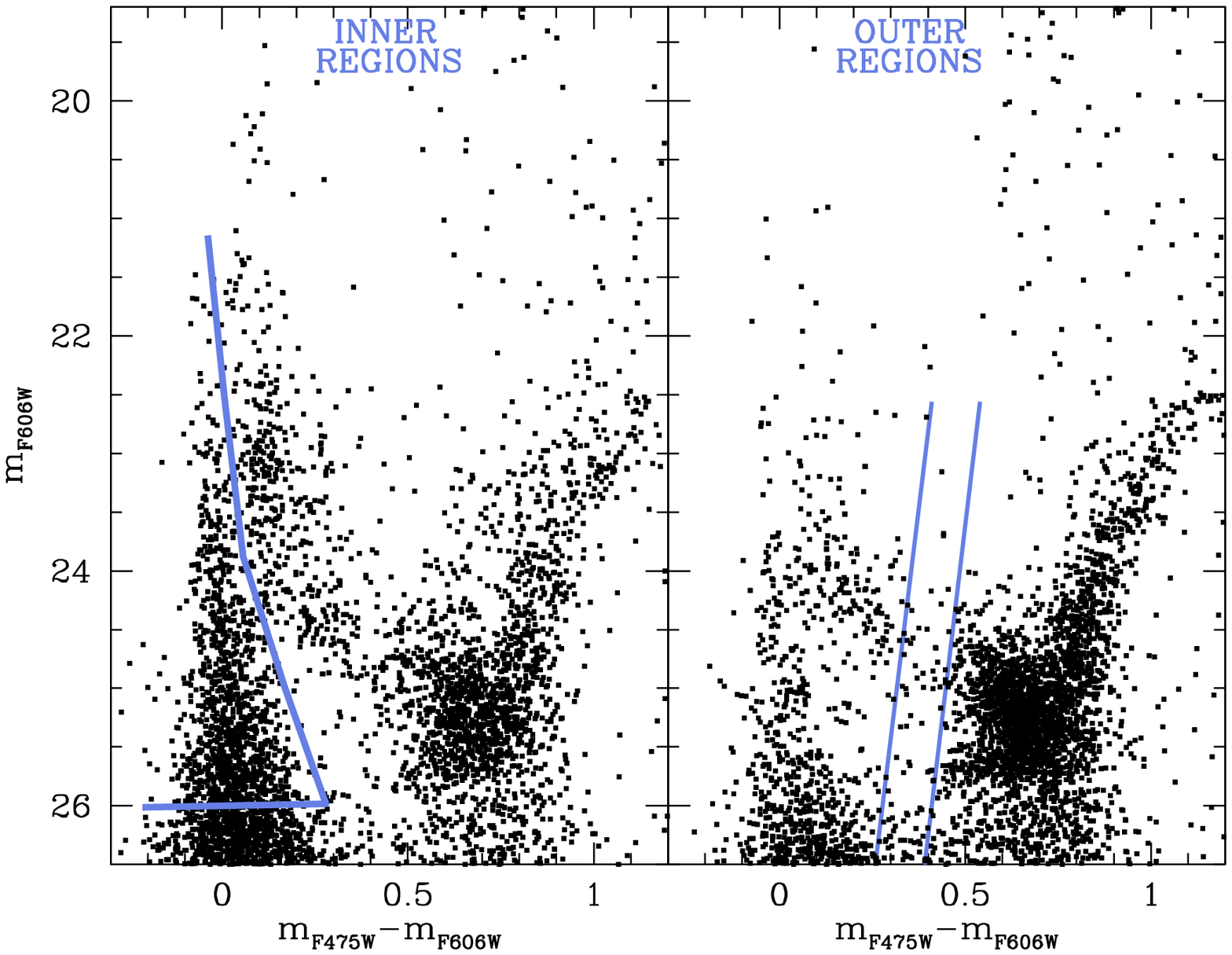}
\caption{Enlargements of the color-magnitude diagram illustrating the
stellar population gradient in SagDIG.  {\it Left:}\  The CMD of stars
within the star forming regions, near the  center of SagDIG.  The {\it
curved  line} delimits the selection  of MS  stars used in calculating
the star formation  rate (see Sect.  5.2).  Note  the presence  of the
extended  blue-loop  feature and  its red  counterpart parallel to the
RGB, indicating   a significant population  younger than  1 Gyr.  This
young population  is less prominent  in the  {\it right} panel showing
the  \vv, (\bbvv) diagram for stars  outside the star forming regions.
The richness of the red clump and its complex structure are indicative
of  a predominant intermediate-age  population formed over an extended
epoch.  The noticeable,  although minor, extension to lower luminosity
is  reminiscent  of the  red horizontal  branches  seen in  most dwarf
spheroidal  galaxies.   The {\it   solid lines} show   the approximate
location of the instability strip.
}
\label{f_cmdout}
\end{figure*}
%-------------------------------------------------------------

%%%%%%%%%%%%%%%%%%%%%%%%%%%%%%%%%%%%%%%%%%%%%%%%%%%%%%%%%%%%%%%%%%%%%%
%%								    %%
\section{Color-magnitude diagrams}                                  %%
\label{S_CMD}
%%								    %%
%%%%%%%%%%%%%%%%%%%%%%%%%%%%%%%%%%%%%%%%%%%%%%%%%%%%%%%%%%%%%%%%%%%%%%

%----introduction
%
The ACS  color-magnitude diagram of  SagDIG reveals for the first time
its  history    of   star  formation    (Fig.~\ref{f_maincmd}).    The
high-resolution   of   HST/ACS    allows    overcoming  the   crowding
difficulties,  producing  highly detailed  CMDs where many distinctive
features of the various  stellar  populations are  seen for  the first
time.   Figure~\ref{f_maincmd} presents both  the \vv, (\bbvv) and the
\ii, (\bbii)  diagrams,  which complement each  other  in highlighting
different evolutionary phases.  The CMDs presented in this section are
limited to stars with absolute SHARP values below 0.3.
%

%-----Foreground stars:
%
Some foreground  contamination is  present  in all diagrams.   This is
dramatic in the case  of  ground-based observations,  yet it does  not
significantly affect the ACS data. Indeed, because  of the distance of
SagDIG,  most of the galaxy stars  are fainter  than the main-sequence
turnoff(s) of the thick disk stars,  which are the main contributor to
the foreground counts.  In this respect, ACS really  allows us to look
{\it through the Milky Way}.

As an aid to  the interpretation of  our CMDs,  we have simulated  the
color-magnitude diagram of the Galactic stars  along the line of sight
of   SagDIG  using     the   \besancon\  Galaxy     Model  (Robin   et
al.      \cite{robin03};          see          also               {\sf
http://bison.obs-besancon.fr/modele/}).   This  is shown  in the upper
right panel of Fig.~\ref{f_maincmd}.  The foreground stars belong to 3
main   populations: (i) the Galactic  disk,  (ii) the  thick disk, and
(iii) the halo.
Although the simulated diagram  cannot  be employed to  quantitatively
reproduce the observed Galactic contribution, because of the different
pass-bands, the  overall similarity   of  the simulated  and  observed
diagrams is helpful to interpret the observed CMD features.
The vertical sequence  at  (\bbvv)~$=1.4$ is clearly produced  by  the
superposition of  the {\it  lower} main  sequences of  Milky Way stars
projected throughout the line of sight.
The secondary peak located  at ($B-V$)$\simeq0.5$ corresponds  to the
overlapping of  G field main-sequence turnoff stars  along the line of
sight.   For    example,   stars    at   \vv~$\simeq20.0\pm0.2$    and
\bbvv$\simeq0.55$ would  correspond to  the turnoff  area  of Galactic
stars located at $\sim8$ kpc from us.
These stars are quite important on their own, as they provide clues on
the  Galactic structure.  A     thorough analysis  of   the  projected
distribution of Milky Way stars will  be possible at the completion of
our program,   when second-epoch data   will enable us to  analyze the
proper-motion of all the observed populations.

%------Ok: back to sagdig
%
%------RGB
%
Turning again  our attention to  SagDIG, we  note  that the red  giant
branch (\abbrev{RGB}) is now well sampled  from its tip at \ii$\approx
21.3$ down  to \ii$\sim  26$, i.e. well   below the red clump  of core
He-burning stars.
%
%----RSG
This RGB is accompanied  by a parallel and  slightly bluer sequence of
red   super-giants (\abbrev{RSG}), whose   most   luminous  stars  are
confused with the Galactic foreground.
The red   super-giants  are better seen in   Fig.~\ref{f_cmdout} (left
panel), where we plot the CMD of stars in the  inner region of SagDIG.
This  diagram is dominated  by the  young  stellar population.  On the
other  hand,   the CMD  of   the outer regions  (see   right  panel of
Fig.~\ref{f_cmdout})  shows, as   expected, a   decreasing fraction of
young stars.

%------------------ YOUNG POPS. --------
%
One of  the   most noteworthy features    of the  ACS  color-magnitude
diagrams in Fig.~\ref{f_maincmd}    is the perfect separation of   the
H-burning {\it main  sequence}   and the sequence of   blue He-burning
stars (\abbrev{BHeB}) in  the blue-loops evolutionary phase.  This  is
best seen in the \ii, (\bbii) diagram of Fig.~\ref{f_maincmd}. Near
\ii $=23$, the color difference between the two populations reaches
(\bbii)$=0.35$  mag.  Besides the  photometric precision  of ACS, this
result is obviously  due  to the wide   color baseline chosen for  our
observations.
At \ii$\simeq25$, the \abbrev{MS}  shows a clear bifurcation.  A small
component   departs from  the main   sequence  and proceeds obliquely,
crossing the blue loops at \ii$\sim  24.0$. This feature is indicative
of  a    $\sim    200$  Myr  old  burst     of star   formation   (see
Sect.~\ref{S_recentSF}).   A similar  feature  is seen  (although at a
slightly lower resolution because  of the narrower color baseline)  in
Sextans\,A (Dolphin et al. \cite{dolphin03}).
We also notice a gap in luminosity between \ii$=20.6$ and 21.1, in the
distribution of  blue   super-giants centered  at  (\bbii$\simeq0.5$).
Although this hint is based on  a statistically small sample of stars,
we note that   any quiescent star-forming   phase in the past  100 Myr
would not be detectable along the MS, since successively younger stars
would  overlap, and  therefore fill  any {\it empty}  region along the
main sequence.
On the contrary, evolved blue or  red super-giants {\em do not overlap
in luminosity}, therefore   preserving any  signature of a   quiescent
phase.
Thus, the  blue   super-giant gap between  \ii$=20.6$ and    21.1 (at
\bbii$\simeq0.5$)  might trace a {\it   short quiescent phase} that we
estimate (via isochrone fitting)  took place some time between 30$-$60
Myr ago.

%------------------------- I-AGE POPS. --------------

%---C stars
Another important feature  in the CMD of  SagDIG is the presence of an
extended   asymptotic  giant branch    (\abbrev{AGB})  which traces  a
conspicuous intermediate age population.  The AGB  in SagDIG is partly
masked   by    the   foreground   vertical    sequence   centered   at
(\bbvv)\,$=$1.4, and becomes really evident  only when ACS photometric
catalogs are cross-identified with the results of carbon star surveys.
Figure~\ref{f_maincmd} provides a cross-identification of our ACS data
with the carbon star samples of Cook (\cite{cook87}; open squares) and
Demers \& Battinelli (\cite{demers02}; open circles).
This comparison confirms  the composite  nature  of the  SagDIG carbon
stars sample.   As suggested by Cook  (\cite{cook87}), the C  stars in
SagDIG fall  in    two classes:   (1)  a luminous,    intermediate-age
population reaching $\sim2$  mag in $I$ above  the RGB tip;  and (2) a
bluer, low-luminosity component,  overlapping the upper  1 mag  of the
RGB, probably belonging to an older population.
We  mention a  third possibility for  the  origin of  the carbon stars
identified in the SagDIG field. Of the  3 stars that coincide with the
vertical     foreground      sequence     (upper-left       panel   in
Fig.~\ref{f_maincmd}),  the brightest one can  be a luminous object on
the extended AGB.  However,  the  two fainter stars, quite   separated
from all other carbon stars, might belong  to the Galactic foreground.
These  are possibly related to  the population  of faint high-latitude
Carbon stars found in the Sloan Digital Sky Survey (Downes et al.
\cite{downes04}).  The possibility that these two stars are Galactic
low-luminosity dwarf Carbon stars will be better disentangled once our
proper-motion analysis of SagDIG is completed.

%---RC:
%
The color-magnitude diagrams in  Fig.~\ref{f_maincmd} also show a {\it
prominent red clump}  (\abbrev{RC}) of core He  burning stars, with  a
complex,     elongated    shape.  This    structure  (which    will be
quantitatively analyzed along with  the star formation history  of the
galaxy in a separate paper --  Held et al. 2005,  in prep.)  indicates
that the  bulk of SagDIG  stars  formed at  intermediate ages ($\ga 1$
Gyr), in a relatively extended star formation episode.
%
%- 
The extension of the RC at higher luminosities, sometimes known as the
``vertical clump'' (\abbrev{VC}), marks the evolution of younger stars
($\sim0.35-0.63$ Gyr) in the ``blue loop'' core He-burning phase.
%

%------------------ OLD POPS. --------
%------Red HB:
%
In addition to these components, we note a fainter extension at
\vv$=25.70\pm0.05$ (\ii~$\sim25.2$) identified with a red horizontal
branch (\abbrev{HB}).  Again, spatial  selection helped us to pinpoint
specific stellar populations.  The presence of an old red HB is better
seen in Fig.~\ref{f_cmdout} (right panel), where stars located outside
the regions of  active star formation have  been selected.   Detecting
the  presence of  a truly  old population  in  SagDIG  was one of  the
primary goals of this study and, indeed, this red HB implies that {\it
SagDIG hosts an old ($\ga 10$ Gyr) stellar population}.
%
%---- RR Lyrae 
%
Together, Figs.~\ref{f_maincmd} and  ~\ref{f_cmdout} may  also suggest
that a search for  variable stars, in  particular the old ($> 10$ Gyr)
RR\,Lyrae variables, might be  worthwhile,  even though the number  of
RR\,Lyrae stars we expect  from Fig.~\ref{f_cmdout} is  not high.   In
these figures,  the   blue and red edges    of  the instability  strip
converted    from  the    theoretical    HR     diagram  (Chiosi    et
al. \cite{chiosi92}) to the ACS/WFC system,  have been over-plotted on
the CMD.
The ``gaps'' noticed along the HB [(\bbvv)$=0.3$--$0.4$,
\vv$\simeq25.8$] as well as along the ``vertical clump''
[(\bbvv)$=0.35$--$0.45$,    \vv$\simeq24.8$] show  an almost   perfect
coincidence with  the instability strip.   In particular, the red edge
appears to agree very well  with a {\it  truncation} in the red HB and
blue-loop sequence.
Indeed,  short-period variable stars  have  been found  in other dwarf
irregular galaxies (e.g., Dolphin et al. \cite{dolphin02}; Skillman et
al. \cite{skill03}; Baldacci et al.  \cite{bald+04}).
Unfortunately, a similar search  cannot be conducted using our present
observations because of their limited 3  hours baseline.  Selection of
candidate variable   stars   is   likely   to  become possible    once
second-epoch observations are available.

%%%%%%%%%%%%%%%%%%%%%%%%%%%%%%%%%%%%%%%%%%%%%%%%%%%%%%%%%%%%%%%%%%%%%%
%%								    %%
\section{Reddening, metallicity, and distance}                      %%
\label{S_feh}
%%								    %%
%%%%%%%%%%%%%%%%%%%%%%%%%%%%%%%%%%%%%%%%%%%%%%%%%%%%%%%%%%%%%%%%%%%%%%

%-----------------------------------------
\subsection{Reddening}
\label{s_redd}

%------------------------------------------one column figure
\begin{figure}
\centering 
\resizebox{\hsize}{!}{\includegraphics{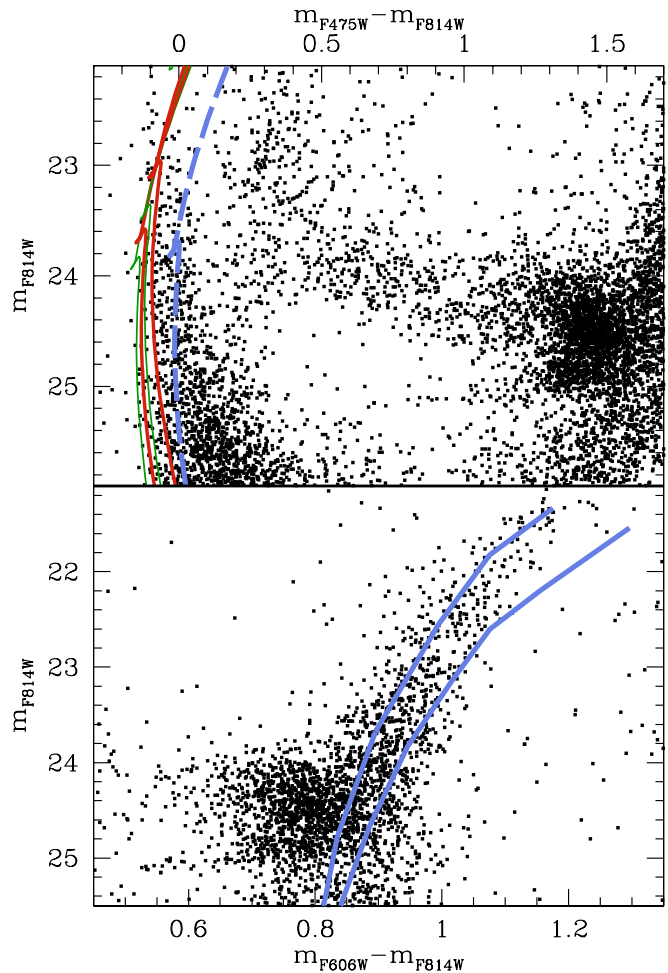}}
%\resizebox{\hsize}{!}{\includegraphics{./psfiles/F_reddening.ps}}
%\centering \includegraphics[width=8cm,height=13cm]{./psfiles/F_reddening.ps}
%
\caption{The {\it upper panel} shows   theoretical isochrones in   the
ACS-Vega  system on the \ii, (\bbii)  diagram.   Solid isochrones have
been shifted using a \dmod~$=25.14$ and \ebv~$ = \ebvsag$, whereas the
long-dashed isochrone has  been shifted using \ebv~$   = 0.20$.  Solid
isochrones, and from left  to  right, are with $Z=0.0001$,   $0.0004$,
$0.001$ and $0.004$ (corresponding to [Fe/H]=$-2.36$, $-1.74$, $-1.33$
and $-0.72$).   The long-dashed  isochrone  is with  $Z=0.0004$.   All
isochrones refer to an age of 63 Myr.
The {\it  lower panel} shows the  \ii, (\vvii) color-magnitude diagram
of  SagDIG with  superimposed    the fiducial lines of    the Galactic
globular clusters NGC\,6752 (right) and NGC\,6341 (left), for which we
used    the  reddening  and distance     moduli  given in  the  Harris
(\cite{harris96})            on-line             catalog:         {\sf
http://physun.physics.mcmaster.ca/$\sim$harris/mwgc.dat},  as  updated
in February 2003.}
\label{f_reddening}
\end{figure}
%-------------------------------------------------------------

The  range  of  \ebv\  values for  SagDIG   in the   literature is not
particularly  large,  spanning the interval between  0.06  (Lee \& Kim
\cite{lee00})   up  to  \ebv~$  =   0.19$,  derived   in  Saviane  et
al.  (\cite{saviane02}) from the Balmer decrement  in an \hii\ region.
Nonetheless, this range   may significantly affect our  conclusions on
the stellar metallicity   in SagDIG.  Momany et  al. (\cite{momany02})
derived \ebv~$ = 0.07$  from the colors of foreground  stars at the MS
turnoff, while the Schlegel et al. (\cite{schlegel98}) extinction maps
indicate  a foreground reddening \ebv~$  = \ebvsag$.  These two values
are likely to bracket the {\it foreground} reddening toward SagDIG.

In addition, some  degree of internal reddening  has been suggested by
Momany et al. (\cite{momany02}) to be associated with the star-forming
regions. Accordingly,  a {single} value of  reddening  for all stellar
populations in SagDIG may not be appropriate.
A comparison   of the ACS   color-magnitude diagrams  with de-reddened
isochrones    appears      to   confirm   this   earlier    suggestion
(Fig.\ref{f_reddening}, upper panel).
In order to  test  the reddening value needed   to fit the young  main
sequence, a set of  63~Myr old isochrones with different metallicities
are over-plotted on the  SagDIG  color-magnitude diagram in  the upper
panel of Fig.~\ref{f_reddening}.
The isochrones  provided  by L.  Girardi  are from  the  Padua stellar
evolution  models,  calculated  in the  same  way  as  Girardi et  al.
(\cite{leo02}), using  the  ACS transformation  curves by Sirianni  et
al. (\cite{siri+04}). These isochrones assume Y$=0.230$.
The conversion from  \ebv\ to the extinction in  the ACS/WFC bands was
taken from Bedin et al.  (\cite{bedin04}).

In the upper panel of Fig.~\ref{f_reddening} we over-plot 4 isochrones
with $Z=0.0001$, $0.0004$, $0.001$ and $0.004$ (respectively from left
to right) shifted  by  \ebv~$=\ebvsag$.  One  sees that  assuming this
reddening  value  even the un-plausible  hypothesis  of metal-rich, 47
Tuc-like, isochrone is  still bluer than the  bulk of the  SagDIG main
sequence.   On the contrary, a reasonable  $Z=0.0004$ (as derived from
the [O/H] of the brightest \hii\ region, Saviane et al.
\cite{saviane02}) isochrone shifted by \ebv~$=0.20$ shows an excellent
fit to the SagDIG main sequence.

A similar comparison with isochrones   cannot be employed for the  RGB
colors    because    of the    well-known   uncertainties      in  the
color-metallicity calibration  of theoretical isochrones for red giant
stars (e.g., Westera et al. \cite{west+02}).   However, we can compare
the data for  SagDIG with the  RGB fiducial lines of Galactic globular
clusters  to check the  foreground reddening of  RGB stars.  The lower
panel of Fig.~\ref{f_reddening} over-plots  the RGB fiducial  lines of
two  Galactic  globular  clusters, NGC\,6341 and   NGC\,6752, onto the
color-magnitude diagram  of SagDIG.  The  clusters were  observed with
the same instrumental setup  (GO 9453, P.I.   Brown; see Brown  et al.
\cite{brown03}) and  re-calibrated by Bedin  et al.   (\cite{bedin04})
with the same methods as used here for SagDIG.
Both NGC\,6341  ([Fe/H]~$=-2.3$) and  NGC\,6752 ([Fe/H]~$=-1.54$) have
been corrected to the   distance  of SagDIG  using the   reddening and
distance moduli given  in  the Harris (\cite{harris96}) catalog,   and
shifted using  \ebv~$= \ebvsag$  and \dmod~$=  \distsag$  to match the
SagDIG diagram.
While this   comparison  does   not provide   itself   an  independent
measurement    of reddening, it gives  no   indication of any internal
reddening outside of star-forming regions.   Therefore in the case  of
RGB stars we adopt the  Schlegel's et al. (\cite{schlegel98}) value as
our estimate of the total reddening.  In light of the results obtained
in M02, we regard this reddening value as an upper limit and associate
to it an  uncertainty  of $\pm0.05$ that  brackets  the  low reddening
value adopted in M02.

%------------------------------------------one column figure
\begin{figure}%[th]
\centering 
%\resizebox{\hsize}{!}{\includegraphics{./psfiles/F_FeH.ps}}
%\includegraphics[width=8.5cm]{./psfiles/F_FeH.ps}
\includegraphics[width=8.5cm]{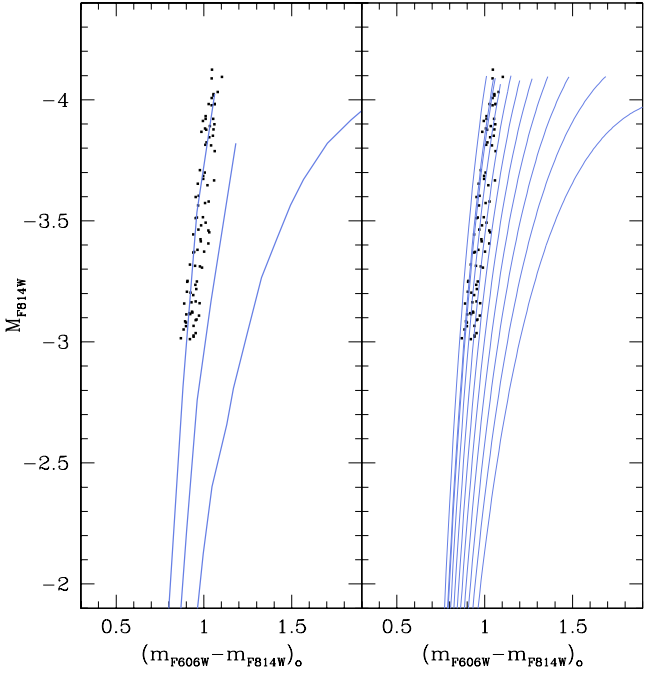}
\caption{
The $2-\sigma$ selected sample of RGB  stars in SagDIG, plotted on the
$M_{\rm F814W}$ $vs.$ (\vvii)$_{\circ}$ diagram.
On the left panel, the data are shown along with RGB fiducial lines of
the Galactic globular clusters NGC\,6341, NGC\,6752, NGC\,104.  A grid
of analytic RGB  templates  are superposed to  the  data in  the right
panel (see text for details).}
\label{f_feh}
\end{figure}
%-------------------------------------------------------------

%-----------------------------------------
\subsection{Metallicity of the red giant stars}

The mean  stellar metallicity of  SagDIG was estimated by  analyzing a
sample of red giant stars selected in regions outside the star-forming
complexes. This was done  using the Saviane et al.  (\cite{saviane00})
method, extended here to the F606W and F814W ACS pass-bands.
The method is based on  the construction of  a family of hyperbolae in
the    plane $M_{\rm F814W}$,   (\vvii)$_{\circ}$  suitable to fit the
observed RGB  fiducial lines of template globular  clusters  in a wide
range of metallicities.
The present implementation   makes use of  the  empirical RGB fiducial
lines of  3  Galactic globular    cluster (NGC\,104, NGC\,6752,    and
NGC\,6341) observed with   ACS.   This set  of globular  clusters  was
originally observed by Brown et  al. (\cite{brown03}) to constrain the
metallicity   of   the  M\,31  halo,    and  reanalyzed  by  Bedin  et
al. (\cite{bedin04}).   The adopted metallicity of  the  3 clusters is
[Fe/H]~=$-$0.76, $-$1.56  and $-$2.28, respectively,  as given  in the
Harris  (\cite{harris96})  catalog.   The   fiducial lines have   been
transformed  into  the $M_{\rm   F814W}$, (\vvii)$_{\circ}$   plane by
adopting (1)   $(m-M)_V   =   13.37$, 13.13,    and 14.64;    and  (2)
\ebv~$=0.04$, 0.04, and 0.02 respectively.

The  application  of this method  to   SagDIG in  the $M_{\rm F814W}$,
(\vvii)$_{\circ}$ plane is illustrated  in Fig.~\ref{f_feh}, where the
absolute magnitude and  reddening-corrected colors of individual stars
in SagDIG are  compared with  the RGB ridge   lines of the  3 Galactic
clusters  (left panel), and  the  grid  of  analytic RGB  lines (right
panel).  The observed magnitude  and colors have been transformed into
the     $M_{\rm F814W}$,    (\vvii)$_{\circ}$    plane  by    assuming
\ebv~$=\ebvsag$ and \dmod~$=25.14$ (see below).
The deepth of our ACS  imaging allows a  much better identification of
the   RGB with  respect    to ground-based   observations. To  prevent
inclusion of young red super-giants contaminating the blue side of the
RGB, we  applied a $2\sigma$ rejection in  color about a fiducial line
drawn by eye.

%- mean metallicity, results:
%
An interpolation over color  provides a metallicity estimate  for each
star.   The  [Fe/H] values   of  the  selected  RGB  sample  span from
[Fe/H]~$= -2.6$ to $-0.8$.  The resulting  mean metallicity (for stars
brighter than $M_{\rm F814W} = -3.0$) is $\langle[\rm{Fe/H}]\rangle =
\fehsag$ with an r.m.s. of $\sigma_{\rm [Fe/H]} = 0.18$. By repeating
the calculation with a lower reddening, \ebv~$=0.07$ (as in M02) we
would obtain $\langle[\rm{Fe/H}]\rangle = -1.89 \pm 0.17$.  The value
measured by Momany et al. (\cite{momany02}) was
$\langle[\rm{Fe/H}]\rangle=-2.10$.  Given the uncertainties and the
different pass-bands and calibrations, we regard these results as
perfectly consistent. According to the new [Fe/H] measurement, SagDIG
is confirmed to be very metal-poor yet ``normal'' among dwarf
irregular galaxies.
We note that this metallicity estimate  does not take into account the
fact that the RGB stars  in SagDIG are younger,  on average, than  the
red giant   stars in globular  clusters. However,  the effect is quite
small for metal-poor stars (see Rizzi et al. \cite{rizz+03}).  We will
implement this    correction  after  discussing the    SFH   of SagDIG
(E.~V. Held et al. 2005, in prep.).

%------------------------------------------one column figure
\begin{figure}
\centering 
\resizebox{\hsize}{!}{\includegraphics{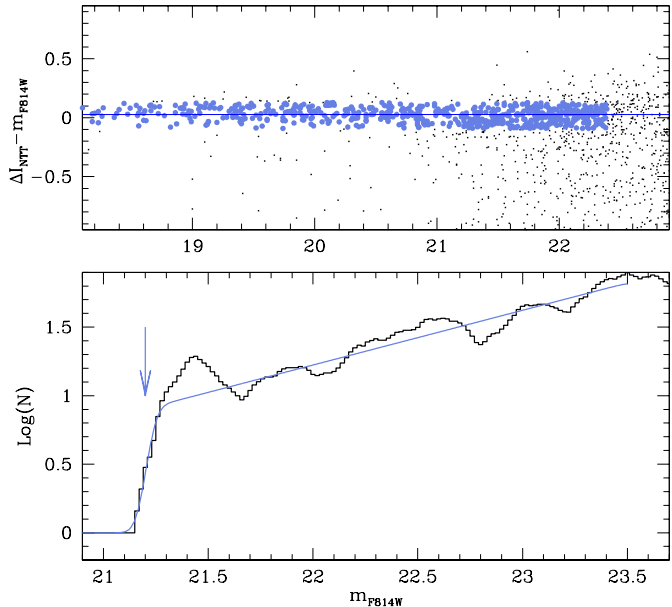}}
%\resizebox{\hsize}{!}{\includegraphics{./psfiles/F_tip.ps}}
%\includegraphics[width=8cm,height=8cm]{./psfiles/F_tip.ps}
%
\caption{The upper panel shows a comparison between the ACS \ii\ 
data    and $I_{\rm  NTT}$       photometry   from Momany  et      al.
(\cite{momany02}).  Stars    plotted  with   heavy    symbols  are   a
$3\sigma$-clipped subsample yielding a  mean shift of $-0.03 \pm 0.03$
mag.
The lower panel  shows the  luminosity function  of SagDIG RGB  sample
selected outside the  star-forming regions.  The thin line  represents
the best fit with an error-convolved step function used to measure the
RGB cutoff.  An {\it arrow} marks the RGB tip at
\ii(RGBT)~$=21.23\pm0.05$.}
\label{f_tip}
\end{figure}
%-------------------------------------------------------------

%-------------------------------------------
\subsection{Distance}

The excellent match between the RGB tip of SagDIG and that of
\abbrev{GGCs} obtained by assuming the distance derived by M02
suggests that the  new ACS photometry  is consistent with our previous
distance estimate.
Given the superior  RGB obtained from ACS  data, it appears worthwhile
to verify the  distance to SagDIG based   on the magnitude of the  RGB
tip.
Figure~\ref{f_tip}  shows the luminosity   function  of RGB stars   in
SagDIG together with our measurement of the  tip. The RGB selection is
the same as used for measuring the metallicity.
The  RGB  tip  was  found at \ii(RGBT)~$=21.23\pm0.05$   by fitting an
error-convolved step function as in M02.  Since the color term for the
filter F814W is small in the color range of RGB stars (see Sirianni et
al.  \cite{siri+04}), we  have been  able  to compare   the new  F814W
magnitudes with  the  ground-based  $I$-band photometry  of M02.   The
result is shown in Fig.~\ref{f_tip}, upper panel.
The median  difference is  very small,   $I_{\rm NTT}-m_{\rm F814W}  =
+0.04\pm0.03$.  By applying   this correction, we obtain $I_{\rm  tip}
{\rm (ACS)} = 21.27 \pm 0.05$. This is only 0.03 mag brighter than the
value  measured in  Momany  et  al.   (\cite{momany02}), $I_{RGBT}   =
21.30$.
%

%---------------------------------->
\begin{figure*}[ht]
\centering 
%\resizebox{\hsize}{!}{\includegraphics{./psfiles/F_HBs.ps}}
%\includegraphics[width=14cm]{./psfiles/F_HBs.ps}
\includegraphics[width=14cm]{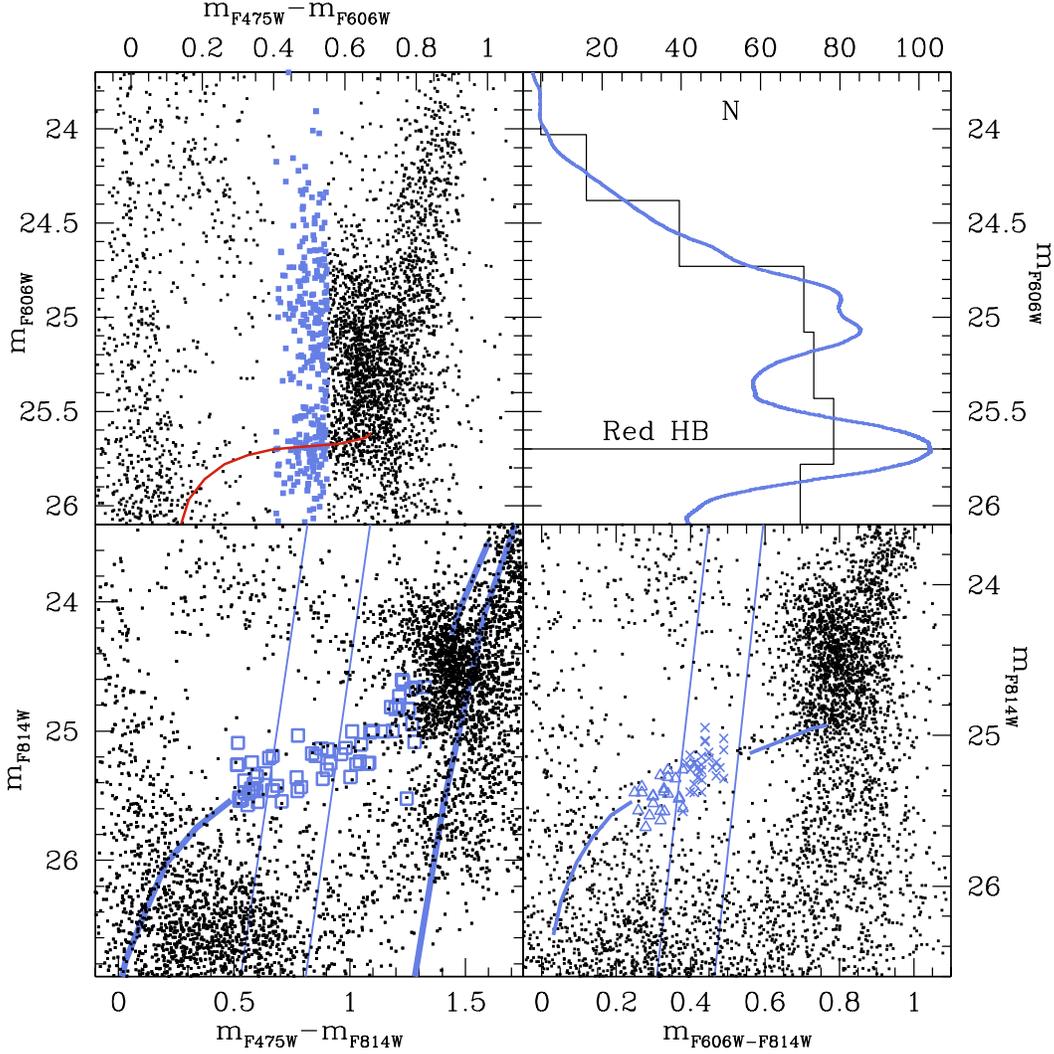}
\caption{Enlargements of the CMDs illustrating the detection of an old
horizontal branch in SagDIG. The upper-left panel highlights the color
interval used  for measuring the  magnitude distribution of He-burning
stars, along with a [Fe/H]$\simeq-2.1$ ZAHB model (assuming Y$=0.245$)
from Pietrinferni et al.  (\cite{piet04}). In the upper-right panel we
plot the histogram and   a multi-bin, smoothed luminosity  function of
stars  selected in upper-left panel.  A line  at \vv~$\sim26.70$ marks
the level of the old red-HB.
In the lower-left  panel, we have  superimposed on the SagDIG data the
fiducial loci for the globular  cluster NGC\,2419 (Momany et al. 2005,
in prep.). These data were fully calibrated in the same way as for the
SagDIG stars.   The HB stars in NGC\,2419  in the  color interval $0.4
<$~\bbii~$< 1.4$ are shown as {\it open squares}.
The Brown et al. (\cite{brown04}) data for  RR Lyrae variable stars in
M\,31 are shown  in the lower-Right  panel ({\it crosses}:  RRab; {\it
triangles}: RRc). Also   plotted is the  fiducial line  for  static HB
stars in M\,31. The Brown's et al.  data have been arbitrarily shifted
to match the red-HB in SagDIG.}
\label{f_hbs}
\end{figure*}
%-------------------------------------------------------------

Assuming the new tip determination at $I=21.27\pm0.05$ and a reddening
of \ebv~$  =   0.12\pm0.05$ (i.e.  $A_I=0.21$  comprising   almost all
reddening  values reported  in  the literature)  and  adopting M$_{\rm
RGBT}^I=-4.04\pm0.05$ for metal-poor systems like SagDIG (see Da Costa
\& Armandroff \cite{dacosta90}, Lee et al. \cite{lee93}, Bellazzini et
al.  \cite{michele01}), the resulting distance modulus as derived from
the calibration of the RGB tip  is \dmod~$=25.10\pm0.11$. The distance
error   includes all uncertainties  on  the  RGB tip determination and
absolute calibration and reddening through standard error propagation.
This basically   confirms    our  previous  M02   distance   estimate,
\dmod~$=25.14\pm0.18$,   $\simeq1.1$  Mpc.  The $0.04$  smaller
distance modulus obtained using the ACS data is to  be attributed to a
slightly brighter    (0.03 mag) RGB   tip   determination, the assumed
M$_{\rm   RGBT}^I$, and  0.05   higher \ebv.    We  note however, that
calibrations of the absolute $I$ magnitude of  the RGB tip are done in
standard ground-based $I$   Cousins filter.  Since  the transformation
from the ACS magnitudes to standard $I$  is quite uncertain, we retain
our ground-based distance determination of SagDIG, and this is adopted
for the rest of the paper.

%----------------------------------------------
\subsection{The Horizontal Branch}
\label{S_HB}

Using  this  distance,  Fig.~\ref{f_hbs} provides  convincing evidence
that the fainter extension of the RC identified in  SagDIG is indeed a
red horizontal branch. In order to estimate the magnitude level of the
horizontal branch,  we extract from the CMD  stars in  the color range
(\bbvv)~$=0.40-0.55$.  The  magnitude  distribution of   these (mostly
core He-burning)   stars  is  shown in  Fig.~\ref{f_hbs}  (upper-right
panel)  as a histogram with  bin size 0.5  mag. A multi-bin luminosity
function is  plotted to reduce the effects  of the choice of the bins.
The red horizontal branch is clearly detected at \vv~$=25.70\pm0.05$.
This is  confirmed  when we   over-plot,  in the upper-left  panel,  a
Zero-Age  HB (ZAHB) model of  [Fe/H]$\simeq-2.1$  from Pietrinferni et
al.    (\cite{piet04}),   having     assumed  \ebv~$=\ebvsag$      and
\dmod~$=25.14$, that  perfectly fit   the estimated  level  of HB   in
SagDIG.

Further evidence for the correct identification of the  HB is given in
the   lower  panels of   Fig.~\ref{f_hbs},  where  we   compare the HB
morphology in SagDIG  with (i) the HB of  an  old, metal-poor Galactic
globular cluster, namely NGC\,2419,  and (ii) the ACS observations  of
variable stars in M\,31 (Brown et al. \cite{brown04}).

NGC\,2419 is one  of the few  clusters matching the SagDIG metallicity
and observed in the same  ACS filters (GO~9666, P.I. Gilliland).  Also
plotted is  a sketch of  the instability  strip.  For the  HB stars in
NGC\,2419 bluer than \bbii$=0.5$ we  only plot a fiducial line because
of the overlap  with the young MS stars   in SagDIG.  This  comparison
shows the consistency of the location the HB level in SagDIG with that
observed in the oldest globular clusters.  On the other hand, it lends
support to the adopted distance modulus and reddening.

In addition,  in Fig.~\ref{f_hbs}   our data  are  compared with   ACS
observations   of  RR      Lyrae    stars  in   M\,31       (Brown  et
al. \cite{brown04}). In this   case, we have arbitrarily adjusted  the
M\,31 data to match  the  red-HB of  SagDIG. This  comparison suggests
where one should search for RR Lyrae variable stars.

%%%%%%%%%%%%%%%%%%%%%%%%%%%%%%%%%%%%%%%%%%%%%%%%%%%%%%%%%%%%%%%%%%%%%%
\section{The recent star formation}	                   
\label{S_recentSF}
%%%%%%%%%%%%%%%%%%%%%%%%%%%%%%%%%%%%%%%%%%%%%%%%%%%%%%%%%%%%%%%%%%%%%%

%-----------------------------------------------
\subsection{Blue loop stars: age and metallicity}

\begin{figure}
\centering 
\resizebox{\hsize}{!}{\includegraphics{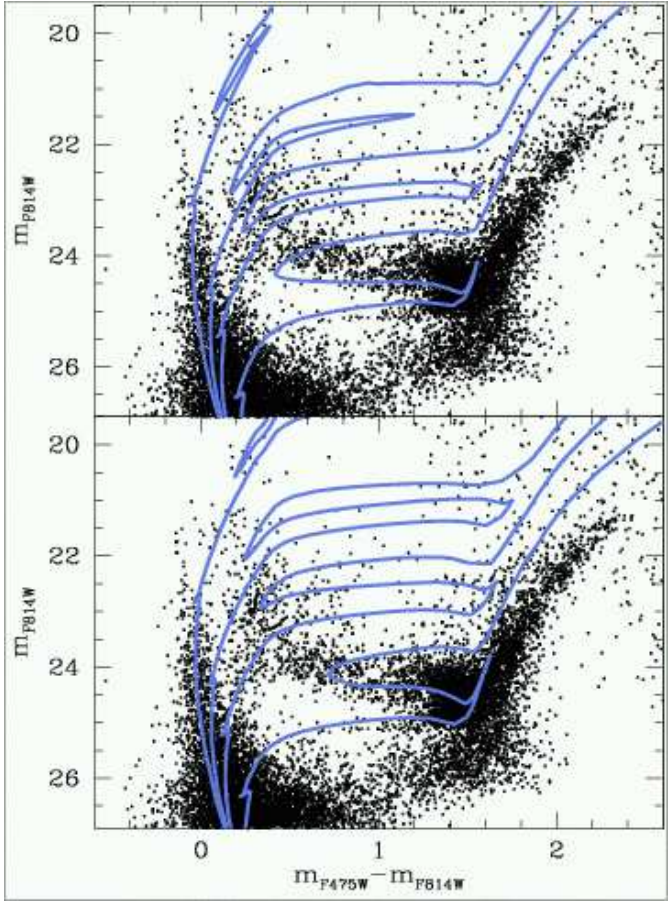}}
%\resizebox{\hsize}{!}{\includegraphics{./psfiles/F_ISOCH.ps}}
%\includegraphics[width=8cm,height=15cm]{./psfiles/F_ISOCH.ps}
%
\caption{Model isochrones from Girardi et al.  superimposed  on the 
\ii,  (\bbii)   diagram of SagDIG.  {\it   Upper} panel: $Z=0.0001$ 
isochrones with  ages of 31.5, 100,  200, 630 Myr.  {\it Lower} panel:
$Z=0.0004$ isochrones with ages 31.5, 89, 200, 630 Myr.  }
\label{f_isoc}
\end{figure}

Theoretical isochrones can now be used to characterize the ages of the
\abbrev{MS} and blue \abbrev{HeB} stars.  Since the metallicity of
SagDIG is known to be in the metal-poor regime, we employed isochrones
with  $Z=0.0001$    and    $0.0004$   ([Fe/H]=$-2.31$   and   $-1.73$,
respectively) for our comparison with observational data.
Very metal-poor isochrones ($Z=0.0001$) with ages  31.5, 100, 200, and
630  Myr are  over-plotted  on the \ii, (\bbii)   CMD of SagDIG in the
upper panel of Fig.~\ref{f_isoc}, while isochrones with $Z=0.0004$ and
ages of 31.5, 89, 200 and 630 Myr are shown in the lower panel.
The MS and  blue  loops are  quite  well reproduced  by both  sets  of
isochrones  from   the  brightest   stars (\ii$=20.5$)  down   to  the
``vertical clump''. However,  the excursion of the  blue loops in  the
isochrones, which is strongly correlated to metallicity, favors a less
extreme metal-poor composition ($Z=0.0004$, or [Fe/H]~$\sim -1.7$) for
the 100--600 Myr old stars in SagDIG.
The details  of  stellar evolution  of  \abbrev{HeB} intermediate-mass
stars depend on many uncertain physical parameters (such as the amount
of   convective overshooting,  stellar rotation,  and   mass loss), so
evolutionary tracks  cannot be    safely used  to  {\it  measure}  the
metallicity of  young HeB stars.  However,  the Padua model isochrones
are known to provide a good fit to systems with known age.  Therefore,
assuming that metallicity is the main driver of  the morphology of the
blue loops (but having in mind all the systematic uncertainties on the
model physics), this  comparison    would suggest a   modest  chemical
enrichment  ($\sim 0.4$ dex)  in stars  since  the main star formation
episode from which most of the RGB stars originated.

%------------------------------------------one column figure
%-------------------------------------------------------------
%: define x (( ((2arcmin/60)/180)*pi*1.07Mpc)*1000000) echo $x
%           622.5007665 pc 
%-------------------------------------------------------------------

\subsection{Star formation rate from main sequence stars}
\label{S_sfr}

%------------------------------------------one column figure
\begin{figure}[ht]
\centering 
\resizebox{\hsize}{!}{\includegraphics{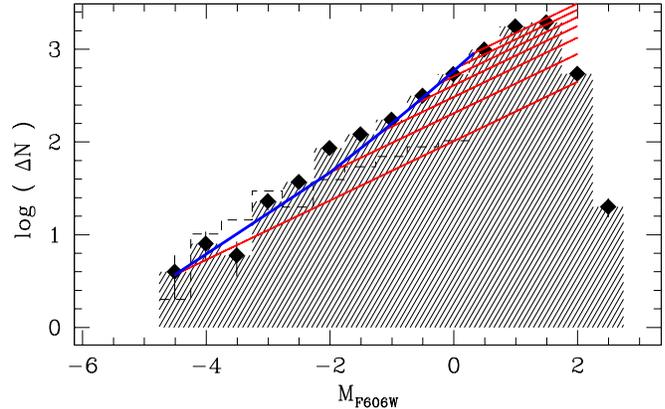}}
%\resizebox{\hsize}{!}{\includegraphics{./psfiles/F_ivo2.ps}}
%
\caption{The $M_{F606W}$ luminosity distribution of MS stars in SagDIG
(shaded histogram).  The {\it   thick solid line} is our  best-fitting
model, which  can   be regarded  as  the  superposition  of luminosity
functions of many SSPs ({\it thin solid  lines}).  The LF of the young
LMC cluster NGC\,2004 is shown as a dashed histogram.}
\label{f_mslf}
\end{figure}
%-------------------------------------------------------------

%--------
While a full analysis of the star formation history of SagDIG based on
synthetic  color-magnitude  diagrams  will  be presented elsewhere, we
give   here  a  first   account  of   the   mean star   formation rate
(\abbrev{SFR}) in the last 0.6 Gyr.
The  SFR   was  computed   as   in  Saviane   et
al.  (\cite{antennae}; hereafter SHR04), using the equation: 
\begin{equation}
\log\dot{n}=(\alpha_{{\rm CSP}}-\alpha_{{\rm SSP}})\,\Delta M_{V}-\log\Delta t
\label{e_ivo}
\end{equation}
\noindent
where $\alpha_{{\rm SSP}}$ and $\alpha_{{\rm  CSP}}$ are the slopes of
the luminosity function (\abbrev{LF})  of a simple (isochronic)  and a
composite stellar population, respectively. $\Delta M_{V}$ and $\Delta
t$ are the  luminosity and  age  interval over which  the  mean SFR is
calculated.
All  the parameters in   the equation  are  measured  on the $V$  main
sequence   LF, except   for $\Delta  t$   which  is obtained  from the
theoretical isochrones.

The main-sequence LF of SagDIG  was obtained by selecting  \abbrev{MS}
stars as indicated  in Fig.~\ref{f_cmdout}, and then constructing  the
$M_V$ histogram with a bin size of $0.5$ mag (Fig.~\ref{f_mslf})
\footnote{We converted the $m_{\rm F606W}$ photometry of MS stars into $V$
magnitude by computing  a mean difference  $\langle V_{\rm NTT}-m_{\rm
F606W}\rangle = +0.05$  mag  (Momany et al.   \cite{momany02}).   This
procedure is acceptable for MS stars since we are considering a narrow
color range around $(B-V)\simeq 0$, while in general the mismatch with
the Johnson  $V$   band    is too large    to   allow   a   meaningful
transformation.  For example, for the  Galactic foreground sequence at
(\bbvv)~$\simeq1.40$,   $\langle V_{\rm  NTT}-m_{\rm  F606W}\rangle  =
-0.50$.}.
The absolute magnitudes were computed as usual from the relation
$M_{V} = V - $~\dmod~$ - R_V $\,\ebv,
adopting a    corrected distance modulus   \dmod~$=\distsag\pm0.18$, a
reddening  \ebv~$=0.19\pm0.04$  (appropriate for    the MS  stars, see
Sect.~\ref{s_redd}), and a ratio $A_V / $\ebv$ = R_{V} = 3.1$.
The    mean     slope of     the    LF    of   SagDIG,   $\alpha_{{\rm
CSP}}=0.486\pm0.022$, was calculated by fitting a straight line to the
LF for $V\leq26$. This  ensures that stars on  the sub-giant branch do
not contaminate  the  MS  sample.  Given  this  limit,  the useful  LF
interval is $\Delta M_{V}=5$ mag.
In order  to  calculate the corresponding  age  interval, we used  the
interpolation of the  MS termination against  age given  in SHR04. The
luminosity range of the LF corresponds  to age limits from $676.7$ Myr
ago to $20.4$ Myr ago, implying $\Delta t=656.2$ Myr.
A   slope  $\alpha_{{\rm SSP}}=0.32$  was  adopted  for simple stellar
populations, as in SHR04.

According        to     Eq.~\ref{e_ivo},       we      find  a     SFR
$\dot{n}=1.032\times10^{-8}\,{\rm  SSP\,{\rm yr}^{-1}}$, normalized to
a  simple  stellar population.   This means that   in  the quoted time
interval ($\sim0.6$ Gyr), $6.8$ SSPs were created.
Finally,  based on the  zero-point  difference between  the  LF of our
template (the young  LMC cluster NGC\,2004)  and a single SSP, we find
that   one  SSP  here  contains $1.32$   clusters  like  NGC\,2004, or
$2.65\times10^{4}\,{\rm M_{\odot}}$  (see SHR04 for further details on
the method).
Therefore  the  mean star-formation   rate in the  quoted interval  is
$2.65\times10^{4}\times\dot{n}=2.73\times10^{-4}\,{\rm M_{\odot}\,{\rm
yr^{-1}}}$.

%- 
Karachentsev et al. (\cite{kara99}; KAM99)  computed the SFR of SagDIG
through a simulation   of the  observed color-magnitude diagram,   and
found        SFR=$13\pm5\times10^{-4}{\rm\,        M_{\odot}yr^{-1}}$,
$6.6\pm0.8\times10^{-4}{\rm     \,         M_{\odot}yr^{-1}}$,     and
$1.3\pm0.1\times10^{-4}{\rm \,  M_{\odot}yr^{-1}}$  in the age  ranges
$0-0.05$~Gyr, $0.05-0.2$~Gyr, and $0.2-15$~Gyr, respectively. In order
to compare  these values with  our rate, we computed the time-weighted
SFR  of  KAM99 within our  time   interval  ($0-0.677$~Gyr), which  is
$3.3\times10^{-4}\,{\rm M_{\odot}yr^{-1}}$, or $\sim20\%$ larger  than
the SFR computed above.
%------------------------
%%
Now KAM99  point out that the  age range $0.2-1$~Gyr is poorly sampled
by their CMD,  so our value  is perfectly compatible with their
SFRs.  A further explanation  for the reported difference is that
KAM99 computed their SFR over  an area  which  is larger than our  ACS
field.

If we restrict our SFR computation  to the $0-0.24$~Gyr interval (i.e.
cutting the LF at $V=24.5$), we obtain  a rate similar to the previous
one ($2.6\times10^{-4}\,{\rm   M_{\odot}yr^{-1}}$),  while KAM99    find
$6.6\times10^{-4}\,{\rm M_{\odot}yr^{-1}}$.
One possibility is that the larger SFR of KAM99 is due to an incorrect
evaluation of  the MS/blue-loop ratio.  This would not  be surprising,
given the large errors affecting  ground-based photometry, that do not
allow a clear separation between the two sequences (as  it is the case
in the ACS data).
We also note that KAM99, using  the H$\alpha$ flux measured by Strobel
et al.  (1991),  i.e. $28.1\times10^{-15}\,{\rm erg\, cm^{-2}s^{-1}}$,
and  applying  the method of Aparicio  et  al.  (2000), find a current
SFR=$1.5\times10^{-4}{\rm  M_{\odot}yr^{-1}}$, much closer  to that
which  we find here.   
In summary,  our method predicts  that the recent SFR remained
almost  constant during the   last few hundreds  of  Myrs. The SFH  of
SagDIG will be further considered with  CMD simulations of the present
ACS data in a future paper.

%%%%%%%%%%%%%%%%%%%%%%%%%%%%%%%%%%%%%%%%%%%%%%%%%%%%%%%%%%%%%%%%%%%%%%
%%								    %%
\section{The spatial distribution of gas and stars}         %%
\label{S_HI}
%%								    %%
%%%%%%%%%%%%%%%%%%%%%%%%%%%%%%%%%%%%%%%%%%%%%%%%%%%%%%%%%%%%%%%%%%%%%%
%
%
The  interplay between stars and   the surrounding interstellar matter
(ISM)  in the form  of neutral, ionized, and  molecular  gas, is a key
aspect in  understanding what triggers star  formation  and drives the
wide range of properties observed in dwarf galaxies.
To investigate this connection,  we analyze here the distributions  of
stars of  different ages in SagDIG  from our  HST/ACS observations and
compare them with the distribution of the ISM.

%-----------------------------------------------------------------
\subsection{The distribution of the gas}
%
%-
%
We begin our discussion by summarizing here the properties of the
\abbrev{ISM} in SagDIG.  The observations of \hi\ in dwarf galaxies
often decompose the neutral H into two components (see, e.g., Young et
al. \cite{young97}, and references therein).
In the particular case of SagDIG, a warm (broader and more turbulent)
\hi\ component having a velocity dispersion $\sigma=10$ \kms\ was
observed,  along with a cold  (narrower and more quiescent) phase with
$\sigma=5$ \kms\ (Young \& Lo \cite{young97}).  The broad component is
distributed all over the galaxy,   whereas the narrower component   is
concentrated in few \hi\  clumps of about $8\times10^5$~M$_{\odot}$ (6
clumps are listed in Table~1 of Young \& Lo \cite{young97}).
A prominent \hi\ clump (with a column density $1.1\times10^{21}$
cm$^{-2}$) is nearly coincident with the biggest, and probably only,
\hii\ region in SagDIG. This \hii\ region is 100 pc across, and 100 pc
away from the \hi\ clump center (Strobel et al. \cite{strobel91}).
These \hi\ clumps  form an almost complete  asymmetric ring (or shell)
structure surrounding a region of lower \hi\ column density, hereafter
referred to as the ``\hi\  hole''.   The roundness  of the outer  \hi\
distribution led Young \& Lo to suggest  an almost face-on orientation
of SagDIG.
The  diameter of  the ring  is nearly  4 arcmin  (1.3  kpc), while the
largest extent of  the \hi\  at a  column density of  $5\times10^{18}$
cm$^{-2}$ is $3.3\times3.0$ kpc.
%

%-------------------------
Young \& Lo (\cite{young97}) found  very little systematic rotation in
SagDIG, i.e.   any rotation velocity is   of the order  of, or smaller
than,  the  intrinsic velocity  dispersion.   However, some  degree of
symmetry in the velocity field led the authors to suggest the presence
of some large-scale organized motion.

%-----
%
Young \& Lo (\cite{young97}) also suggested that the cold dense atomic
phase tells   us  where conditions are    right for the  formation  of
molecular gas and stars.  The sizes of the  \hi\ cold components imply
virial masses which are a few  times larger than  the \hi\ mass alone;
the difference may  be caused by molecular gas.   Thus, the  cold \hi\
component  may   trace  the raw   ingredient necessary  for  any  star
formation activity.
Indeed, it is  found in  clumps  of about the same   size as the  star
formation  regions.    In dwarf irregulars,   star formation typically
occurs in regions of 100-200 pc in size, with each episode lasting for
about  100 Myr and  then  moving  around  (see, e.g., Dohm-Palmer   et
al. \cite{dohm97},\cite{dohm98} for the case of Sextans~A).

%----------------------------------------------------------------->
\begin{figure*}
\centering 
%\resizebox{\hsize}{!}{\includegraphics{./psfiles/F_HI_1.ps}}
%\includegraphics[height=17cm]{./psfiles/F_HI_1.ps}
\includegraphics[height=17cm]{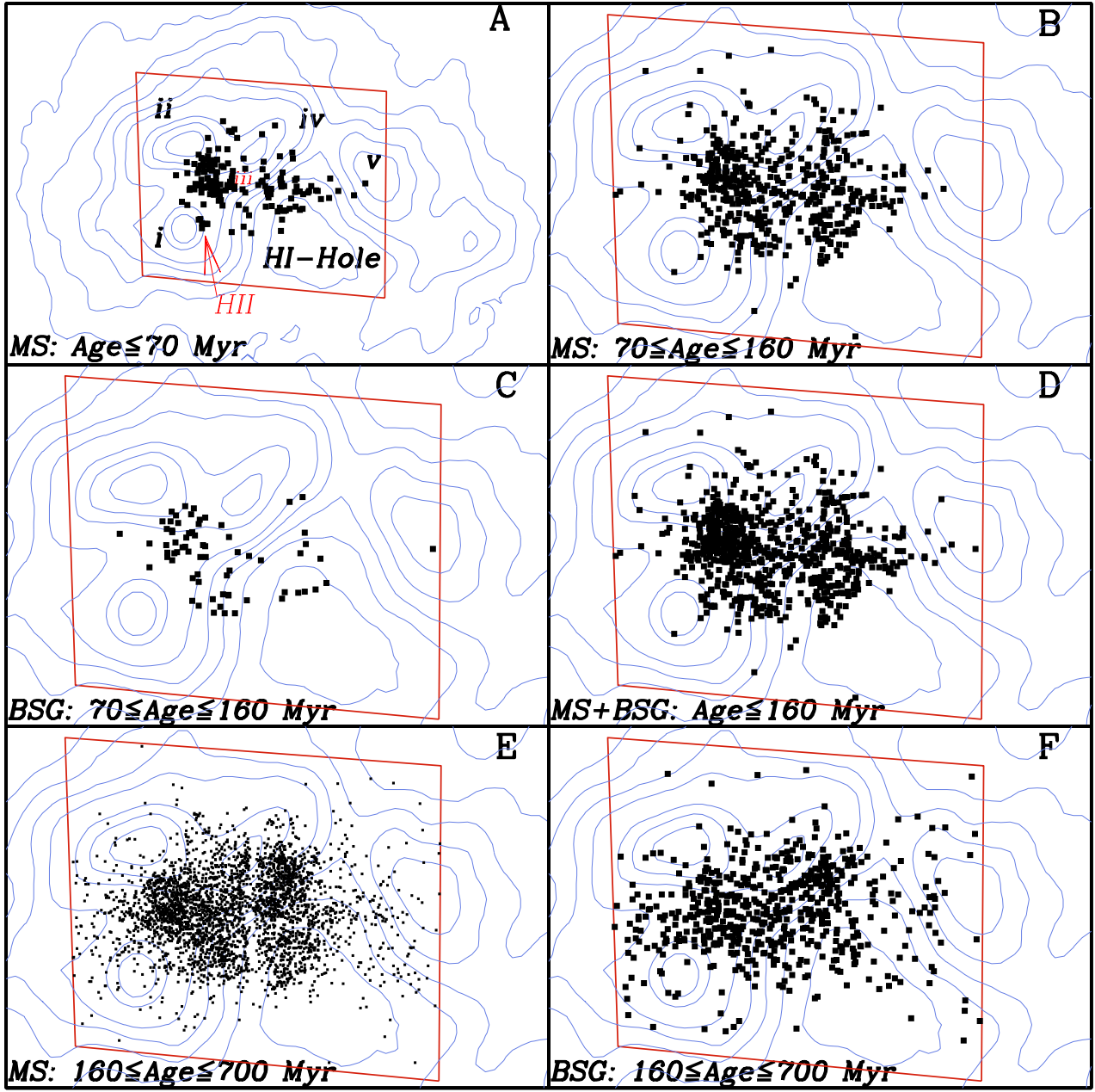}
\caption{ Spatial distribution of MS and BSG stars of different ages
superimposed on the \hi\ contours from Young \& Lo (\cite{young97}).
North is to the top and East to the left. Also plotted is the HST/ACS
footprint.
Panel~A: MS stars with \vv~$\le24.0$ and age less than 70 Myr.
Panel~B: MS stars with $24.0\le$~\vv~$\le25.0$ and age between 70 and
160 Myr.
Panel~C: BSG stars with \vv~$\le23.0$, age 70--160 Myr.  Panel~D: all
MS and BSG stars younger than 160 Myr.
Panel~E: MS stars with $25.0\le$~\vv~$\le26.5$, age interval 160--700
Myr.  Panel~F: BSG stars $22.5\le$~\vv~$\le24.8$, age between 160 and
700 Myr.
}
\label{fig_HI_f1}
\end{figure*}

%-----------------------------------------------------------------
\subsection{Spatial distribution of young and old stellar populations}

The unique photometric deepness and precision  of our ACS data allowed
us  not only  to disentangle stars  in  different evolutionary  phases
(e.g., main-sequence stars   against  blue super-giants) but   also to
assign reliable age estimates to them from a comparison with the Padua
isochrones on the WFC/ACS system (L. Girardi 2004, priv. comm.).
Figure~\ref{fig_HI_f1}  shows the distribution  of  stars in different
age intervals  superimposed  on the \hi\  contour  map of Young  \& Lo
(\cite{young97}), kindly   provided  by  the  authors.     Astrometric
calibration of the radio map and our optical F606W reference image was
performed independently and  carefully  checked.  In all   panels, the
HST/ACS footprint is superposed on   the \hi\ map so   as to mark  the
boundaries of our field of view.
We note  that, in order to  prevent distribution biases  caused by the
spatial variations in the PSF of WFC/ACS, no  selection based on image
shape parameters was applied  to the photometry catalogs when plotting
these   spatial distributions. For  instance,   we found that spurious
distributions may originate from selection  on the $|SHARP|$ parameter
due to the spatially varying image quality.

%--
The location of the youngest  stars in SagDIG, i.e. the  main-sequence
stars with age less than 70 Myr (\vv$\le 24.0$), is plotted in Panel~A
of Fig.~\ref{fig_HI_f1}.  The  MS and BSG stars  were selected  on the
left and right side, respectively, of the  separation line outlined in
Fig.~\ref{f_cmdout}.
The most evident feature is the star-forming region located in between
the \hi\ clumps $i, ii$ and $iii$ on the eastern part  of SagDIG.  The
angular size  of  this  complex  is  $\sim30\arcsec$,  or 155  pc,  in
diameter.  Very recent star formation is also seen in the central part
of the optical galaxy, on the rim of the \hi\ ``hole'', in between the
\hi\ clumps $iv$ and $v$.  A very small complex, or a star cluster,
with only  a few very  massive stars is seen close  to the SE hydrogen
clump ($i$). This  cluster is of  particular interest being associated
with the only \hii\ region in SagDIG.
%-
{Panel B} plots the distribution of MS stars having age between 70 and
160 Myr  ($24.0\le$\vv$\le25.0$).  The stellar  distribution starts to
be more widespread, with a tail extending out to  the \hi\ clump ($v$)
on the western side.  The blue-loop super-giants in the same age range
70--160 Myr  (\vv~$\le23.0$, $0\le$~\bbvv~$\le0.2$), shown in Panel~C,
are consistent with this pattern, although their number is very small.
%
%-
Panel~D  of Fig.~\ref{fig_HI_f1}  summarizes  the  distribution of the
most recent star  formation in  SagDIG  (age $<160$  Myr), showing its
nearly  circular  symmetry,   with a  prominent eastern   star-forming
complex and a peculiar tail of  massive stars reaching the westernmost
hydrogen cloud.

%-
The distribution of older MS  stars ($25.0\le$~\vv~$\le26.5$) and blue
super-giants ($22.5\le$~\vv~$\le24.8$,  \bbvv~$<0.4$; age between  160
and 700 Myr) in shown in Panels~E and F.
The eastern star-forming  region, which  is  main site of very  recent
star formation, is also present for stars older than 160 Myr.  In this
age range, however, a  complex on the  western side of \hi\ cloud $iv$
becomes also prominent.    The latter appears  to have   been the most
active star-forming region more than a few hundred million years ago.
%

%-- a first summary for young stars:
Figure~\ref{fig_HI_f1} gives us a   clear picture of the  relationship
between the \hi\ and   the recent  star  formation ($\le700$   Myr) in
SagDIG.    It confirms  the    earlier    finding  by Young  \&     Lo
(\cite{young97}) that recent   star  formation sites are always   near
(although not coincident) with the \hi\ column density peaks.  This is
best seen for the eastern  and western \abbrev{SF} complexes, as  well
as the small association coincident with the \hii\ region.

%--------------------------------------------------------------->
\begin{figure*}
\centering 
%\resizebox{\hsize}{!}{\includegraphics{./psfiles/F_HI_2.ps}}
%\includegraphics[width=15cm]{./psfiles/F_HI_2.ps}
\includegraphics[width=15cm]{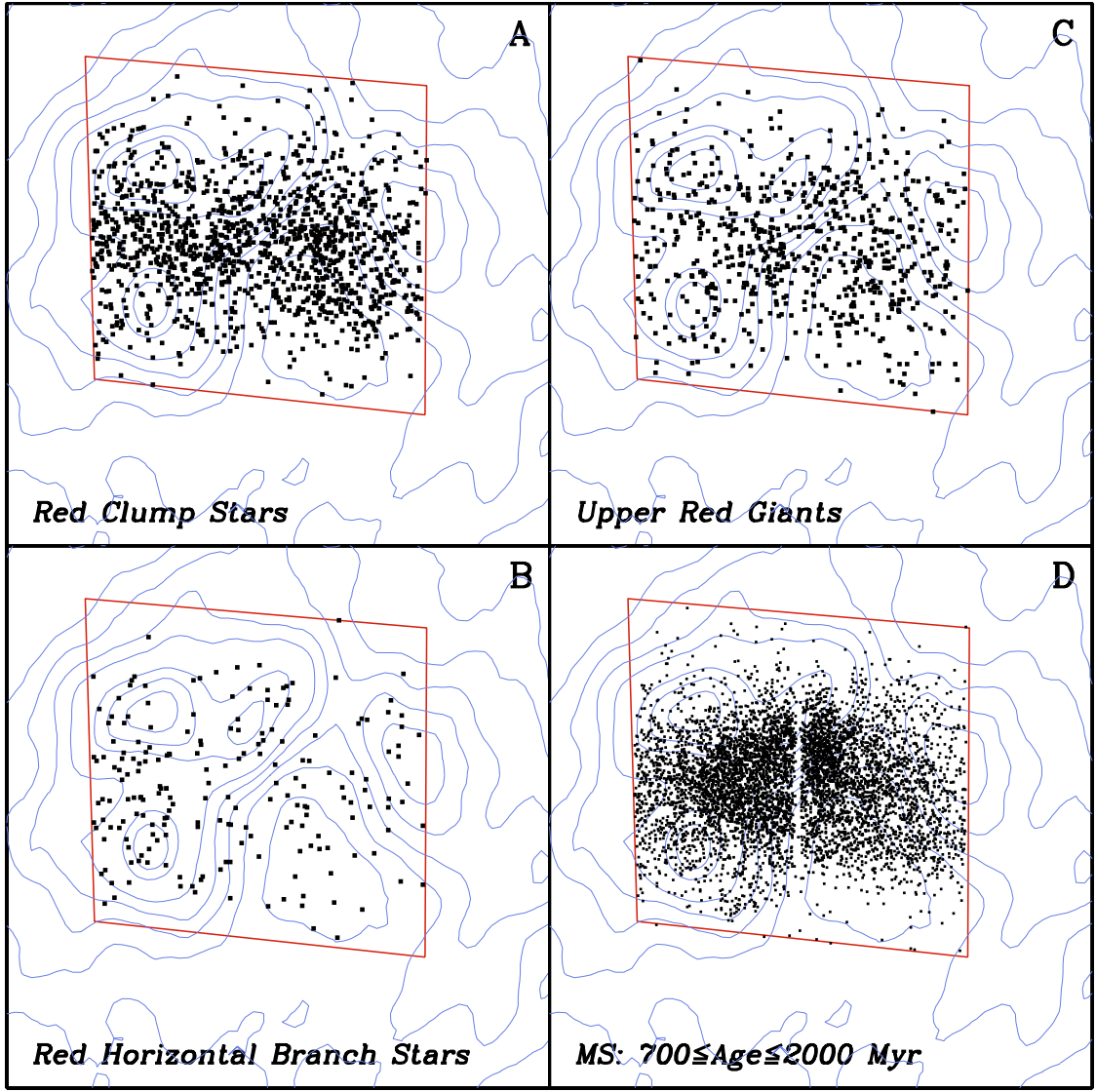}
\caption{Spatial distribution of relatively old stars. 
Panel~A shows the distribution of bright RC stars with
$24.8\le$~\vv~$\le25.4$ and $0.5\le$~(\bbvv)~$\le0.7$,
corresponding to an age interval between 0.3 and 2.0 Gyr.
Panel~B shows the distribution of presumably old red-HB stars, having
$25.6\le$~\vv~$\le25.8$ and $0.4\le$~(\bbvv)~$\le0.6$.
Panel~C shows the distribution of RGB stars brighter than
~\vv~$=24.5$. Red giant stars are at least 1 Gyr old.
For comparison, Panel~D shows the distribution of MS stars with age
between 0.7 and 2.0 Gyr ($26.5\le$~\vv~$\le27.5$).}

\label{fig_HI_f2}
\end{figure*}
%------------------------------------------------------------------|

%-- Fig.12, older stars:

The  spatial  distribution    of relatively    old  stars,   including
main-sequence stars  older than 0.7 Gyr,  He-burning stars on  the red
clump and the red-HB, and RGB stars, is shown in Fig.~\ref{fig_HI_f2}.
Panel~A    shows      the    distribution  of    bright      RC  stars
($24.8\le$~\vv~$\le25.4$, $0.5\le$~\bbvv~$\le0.7$). A  comparison with
isochrones indicates  for these stars an age  interval between 300 Myr
and 2.0 Gyr.
The  distribution  of the  old  red-HB stars ($25.6\le$~\vv~$\le25.8$,
$0.4\le$~\bbvv~$\le0.6$) is shown   in Panel~B. These  stars are  most
likely older than  9-10 Gyr.   A similar plot   is done for  RGB stars
brighter than \vv~$=24.50$ in Panel~C.

For the stars younger than 1 Gyr, a  correlation with the distribution
of the gas and the star-forming regions can still  be noticed. The red
clump stars, for  instance,  appear to follow the   the  rims of  \hi\
clumps on the eastern side. Yet, there is no particular hint of a \hi\
hole imprint.  This  trend is not present  anymore in RGB stars (older
than 1 Gyr) which show some central  concentration.  For the genuinely
old (red-HB)  population the statistics is  really poor --  all we can
say is that their distribution is rather uniform and extended.

It is interesting to compare  these distributions with the location of
MS stars  older than  0.7  Gyr ({$26.5<$~\vv~$<27.50$},  Panel~D). The
fainter limit roughly corresponds to our 50\% incompleteness level, or
an age of about  2.0 Gyr. The  \hi\ hole region is smoothly populated,
and the star distribution is -- already in this age interval -- fairly
uniform with an  overall central concentration,  and  extended.  There
are no detectable voids  in the distribution  of stars.   However, the
distribution is clearly inhomogeneous,  still showing the  presence of
the north-western clump,  and  a generally  higher surface density  of
stars on the eastern  region.  This implies that  the sites  of recent
star   formation discussed for the younger   stars,  in particular the
north-western site, are quite old being already in place 1-2 Gyr ago.

%--------------------------------------------------------------------
\subsection{The \hi\ Hole}           
\label{S_marcel}
\cite{}

If the large \hi\ hole  in SagDIG is  a structure that has formed over
time from an  initially more uniform   distribution we can attempt  to
address the question of how it  formed using constraints from our deep
ACS imaging.    Before we continue however  we  should stress that the
assumption that the hole formed from a  uniform distribution of atomic
gas may be  unfounded.  This point  is discussed  at  the end of  this
section.

A popular scenario for the formation of \hi\ holes is as the result of
the energy input to the ISM from massive stars in  the form of stellar
winds and more   importantly supernovae.  Several galaxies   have been
observed with \hi\  holes, and in  some cases these holes are observed
to be expanding,  although this is generally  observed only  for small
holes (refer to  HoII,  M33, LMC, SMC,  M31, Sextans  A).  Opinion  is
divided, even for a  given galaxy, as  to whether the supernova-driven
wind hypothesis  is   viable or    not  [see for   example  Rhode   et
al. (\cite{rhode99}) and Stewart et al. (\cite{stewart00})].

If the \hi\ hole in SagDIG formed from the combined energetic input of
supernova explosions  we might expect to  detect the light  from those
stars which formed along with those that  drove the expansion but that
have  not exploded as supernovae.  We  would expect  these stars to be
found in a  cluster or clusters that  should still be somewhere within
the \hi\ hole (it is extremely unlikely that  such clusters would have
dispersed  completely during the formation  time of the hole). Because
the stellar  surface   brightness within the  \hi\  hole  in SagDIG is
clearly very low we  will make assumptions in this  estimate so as  to
minimize the stellar luminosity which should remain.

From the \hi\  properties of the \hi\  hole in SagDIG  we can estimate
the energy  required to create it from  the expression given by Heiles
(\cite{heiles79}):

\begin{equation}
E_{\rm E} = 5.3 \times
10^{43}\,n_{0}^{1.12}\,\big(\frac{d}{2}\big)^{3.12}\,\sigma_{\rm v}^{1.4}
\quad \rm erg
\end{equation}

\noindent
where  $d$ is  the diameter of  the hole  in  parsecs, $n_{0}$ is  the
hydrogen particle density in $\rm cm^{-3}$ and $\sigma_{\rm v}$ is the
velocity dispersion of   the atomic gas   (in the  case of  SagDIG  no
expansion  of the \hi\ hole   is detected). This   is a lower limit if
there is expansion in the  plane of the  galaxy which is hidden by the
face-on orientation of SagDIG. The detailed \hi\ observations of Young
et  al. have allowed   all of these to   be measured; they find, $d  =
1.3\;\rm kpc$, $\sigma_{\rm  v} = 6 -  11 \;\rm km\,s^{-1}$ and $n_{0}
\simeq  0.22\;\rm  cm^{-3}$    for   the higher  velocity   dispersion
component. Taking $\sigma_{\rm v}  =  10\;\rm km\,s^{-1}$ we find  for
the hole in the SagDIG $E_{\rm E} \sim 1.5 \times 10^{53}\;\rm erg$.

Assuming that a SN supplies $10^{51}\;\rm erg$ (McCray \& Kafatos 
\cite{mccray87}) of mechanical energy to the  ISM this energy 
requirement implies that it was created  by $\sim 150$ SNe. Assuming a
normal   (Salpeter)   IMF we therefore      normalize the stellar mass
distribution  so that  there are 150   stars more massive than $7\;\rm
M_{\odot}$, which gives an IMF, $\Phi_{m}(m) = 2880\,m^{-2.35}$.

For this IMF the luminosity of all the stars less massive than $7\;\rm
M_{\odot}$ will  depend on  the age  of the  population. Although this
will fall  with time the reduction  in luminosity will  be modest over
the lifetime of the  hole. For the moment we  assume the hole has  the
same age as the lifetime of  the lowest mass progenitor that undergoes
a supernova, $\sim 5 \times 10^{7}\;\rm yr$ (Oey \& Clarke
\cite{oey97}) but see below.

We  estimate a  lower limit to  the luminosity  of the remaining stars
just after the  last supernova occurred by assuming  that they are all
still on the main sequence.  Using a standard main sequence luminosity
function and  integrating from  0.1  to $7\;\rm  M_{\odot}$ we find  a
luminosity of $4500\;\rm L_{\odot}$. In order to reduce this number we
must invoke aging of the population, but stellar evolution models show
that  between $5 \times 10^{7}$ and  $10^{8}$ yr this  amounts to less
than a factor of 2.

At the distance of SagDIG $4500\;\rm L_{\odot}$ corresponds to $m_{\rm
v}\simeq 21$.
Within the \hi\ hole we see no evidence of  any clusters brighter than
$m_{\rm  v}  \simeq 26.0$  and so  if the star  formation episode that
formed the hole  was confined to one or   even several clusters  these
clusters  must have completely dispersed  on  a timescale of $5 \times
10^{7}$ years.
Perhaps  the star formation that formed  the  hole was not confined to
clusters but  was more widespread?   Although the  total stellar light
integrated within  the  \hi\ hole does  exceed  21 mag  this light  is
distributed   smoothly  with  an  intensity  that  increases gradually
towards the optical  center of the galaxy.  The surface  brightness is
always below that  in other regions of the  galaxy.  For the supernova
hypothesis to be feasible then, we require that the stars with ages in
the approximate  range  50--100 Myr are  concentrated within  the \hi\
hole region of  the galaxy.  In Fig.~\ref{fig_HI_f1}  we see that  the
only stars in this age range  within the \hi\ hole  are located at its
NW  edge and that there  are more stars in   this age range beyond the
confines of the hole.  Indeed \emph{there  is no  evidence of  any age
stellar population that  is preferentially distributed within the \hi\
hole}.

Possible ways to reconcile  the  data with the  supernova-driven  wind
hypothesis.

\begin{itemize}

\item The cluster or clusters in which the stars formed have been
completely  dispersed within $\sim  5   \times 10^{7}\;\rm yr$. For  a
stellar cluster   to  be so  short-lived one   would  expect a dynamic
disruption (such   as the  passage through   a  spiral  arm  or  tidal
interaction) to be the  cause. Neither of these  seem to be present in
SagDIG and so one would expect clusters to be long lived.

\item   The hole is very old.  As the \hi\  hole shows no evidence for
expansion it  could have formed  some time in the past  and  just be a
relic    of an ancient star    formation  event. However, the velocity
dispersion of the \hi\  which is observed  to be $10 \;\rm km\,s^{-1}$
places a limit on how  long structure can survive. In  the case of the
hole in SagDIG, with a radius of $650\;\rm pc$ this  is $\sim 6 \times
10^{7}\;\rm yr$.   Therefore, adding the  time of formation, the stars
that formed coevally with those  that expanded the  hole should not be
older than $\sim 10^{8}\;\rm yr$.  Stellar  evolution models show that
for the luminosity of a  cluster to fall by  a factor of 100 (from mag
21 to  our detection limit of  26) requires $\sim 2 \times 10^{9}\;\rm
yr$. We cannot therefore  reconcile our non-detection of clusters with
the presence of the \hi\ hole.

\item  A very  top-heavy IMF. The  analysis  in first part of  this
paper gives no support for anything but a normal IMF in SagDIG.

\item The energy  imparted to the ISM  by a supernova is roughly  2
orders  of  magnitude above the value   we  have used of $10^{51}\;\rm
erg$, or the energy required to produce the hole is lower by a similar
amount.  Both  of these would  require that theories  of supernovae be
substantially revised.
\end{itemize}

In the case of SagDIG  the depth of the ACS  data allow us to conclude
that  the supernova   driven  wind hypothesis   is  an \emph{extremely
unlikely} explanation for the shell structure in SagDIG.
Other authors have come  to similar conclusions  for the  formation of
\hi\    holes (Crosthwaite et  al.   \cite{crosthwaite00}; Deul \& den
Hartog \cite{deul90}; Rhode et al.  \cite{rhode99}) but that in SagDIG
is surely one  of the most secure cases  of a failure of the supernova
wind hypothesis.

There are some possible alternatives to the supernova-driven wind
hypothesis.

\begin{itemize}

\item One possibility is that the \hi\ hole region represents a region
where    the  ISM  is dominated   by    molecular  rather  than atomic
gas. Although the presence of molecular gas may not necessarily result
in a detection of CO emission (CO is a poor tracer of molecular gas in
low metallicity environments,  Taylor et al. \cite{taylor98}, Rubio et
al.  \cite{rubio04}) one would expect to  see some  evidence of recent
star formation associated with the molecular gas. The absence of young
stars in this region therefore argues against this interpretation.

\item Another possibility is that the large gaseous ring is the result
of a dynamical  resonance  in a non-axisymmetric  potential.  For such
rings to form however the gas surface density must exceed the critical
density given by the  Toomre stability criterion. We  cannot calculate
this without  knowledge of the  rotation curve and  so for SagDIG this
cannot be tested. However, the  gas surface density rarely exceeds the
critical density  in  dwarf irregular galaxies  in  which the rotation
curve     has  been   measured     (e.g.   ESO215-G009,  Warren     et
al. \cite{warren04}).

\item A mechanism that may apply to SagDIG and also to dwarfs
containing a   single large  \hi\  ``hole'' such   as the Local  Group
galaxies Sextans A (van Dyk et al. \cite{vandyk98}) and LeoA (Young \&
Lo \cite{young96})  but  probably  not  to  dwarfs like  Ho\,II  which
contain many smaller holes could  simply be that the \hi\ distribution
was never uniform in the first place.  The accretion of an intra-group
\hi\  cloud by  a  low  mass dark matter   halo  (in  which some  star
formation had taken place  at earlier epochs)   would result in  a far
from uniform \hi\ distribution.  The  non-zero angular momentum of the
cloud  with respect to the halo  would naturally lead to the formation
of a ring of    gas. The ring would  form   at the radius where    the
gravitational force balanced the  orbital angular momentum. For atomic
gas that has been recently acquired from an 'external' source there is
little reason to expect any correlation between  the stars and the gas
in   the   resulting system.  Interaction   between   accreted gas and
pre-existing  ISM  could have triggered the  most  recent star forming
event (Fig.~\ref{fig_HI_f1}  panel A).  The  \hi\ structure around the
Phoenix dwarf mapped by Young \& Lo (\cite{young97})  in which 2 large
clouds of \hi\ may be present, neither of which is coincident with the
stellar distribution, is consistent with this scenario.

Rhode et al. (\cite{rhode99}) discuss several alternative explanations
for the formation of large  \hi\ holes in Ho\,II, including  gamma-ray
bursts, impacts of high velocity clouds (HVCs), large scale turbulence
and    ionization by   the  intergalactic   radiation    field.  These
possibilities are all equally applicable  to SagDIG. Apart from noting
that SagDIG is very isolated and that there is no evidence of nearby
\hi\ clouds we can do little to test these ideas.

\end{itemize}

%%

%%%%%%%%%%%%%%%%%%%%%%%%%%%%%%%%%%%%%%%%%%%%%%%%%%%%%%%%%%%%%%%%%%%%%%
%%								    %%
\section{Summary} 				    %%
\label{S_discuss}
%%								    %%
%%%%%%%%%%%%%%%%%%%%%%%%%%%%%%%%%%%%%%%%%%%%%%%%%%%%%%%%%%%%%%%%%%%%%%

%- 

We have presented  a study  of  the stellar populations in  SagDIG  to
unprecedented deepness using observations  obtained with the  Advanced
Camera for Surveys  aboard HST. Our HST/ACS  photometry is deep enough
to go fainter  than the main-sequence turnoff of  halo, thick and thin
disk stars in the Milky Way.  Indeed, using ACS  we can observe SagDIG
looking  through the Galaxy.   Our deep color-magnitude diagrams reach
down to \vv~$\sim 27$, \ii~$\sim 27$, which  allowed us to observe the
stellar populations  of SagDIG to a surprising   degree of detail. The
main results are the following:

\begin{itemize}

\item 
As a result of the photometric precision of ACS and our choice of a
wide color baseline, \bbii, the young main-sequence is traced down to
\vv~$\sim 26$, well separated from the sequence of He burning blue
super-giants on the  blue-loop phase. This allowed  us to estimate the
mean star-formation rate in the last 0.6 Gyr.

\item 
There  is  a well defined  RGB, indicating  that  the dominant stellar
population if   older than $\sim 1$  Gyr.  From the  mean color of RGB
stars    we infer  a    mean   stellar   metallicity  [Fe/H]$=\fehsag$
(uncorrected for age effects).

\item 
Previously known carbon stars are  easily identified on the CMD.  Most
of them belong  to  an extended AGB   typical of  an  intermediate age
population.  However, a few C  stars located with high precision below
the tip of the RGB may belong to an old population.

\item 
The distribution of He  burning stars in the CMD  is indicative  of an
extended period  star formation, spanning  a time  interval between as
early as 10  Gyr  ago  and $\sim 100$   Myr  ago.  Most  of the  stars
populate a very rich ``red clump'', which is the  locus of stars older
than $\sim 1$ Gyr. The stars younger than 1 Gyr are more luminous than
the RC, evolving on the  blue-loop phase to form  the sequence of blue
super-giants,   and the associated   (less  prominent) sequence of red
super-giants, contaminating the blue  side of the RGB.  The luminosity
distribution of the blue super-giants will be used in a separate paper
to reconstruct the history of  star formation up to  1  Gyr ago.   The
wide color range of  blue-loop stars suggests  that even the  recently
formed stars are  substantially metal-poor, among  the most metal-poor
in dwarf irregular galaxies ($Z \approx 0.0004$).

\item 
The presence of a red HB, noticed as a fainter extension  of the RC --
proves that an early epoch of star formation took  place in this dwarf
irregular galaxy. This result is of  foremost importance since it sets
a lower limit to the  first  epoch of  star  formation in yet  another
actively star-forming dwarf galaxy.
%--
In this regard, the resulting scenario is one in which {\em all} Local
Group  galaxies  show to have  had  an  ancient ($\sim10$ Gyr) initial
episode of star formation. The only one exception to this result is
\zwi\  (Izotov \&  Thuan \cite{izo04}) located at  $\sim15Mpc$  and
believed  to be  relatively   young  ($\le2$  Gyr). In  the   {\bf
\it appendix} we briefly comment on this.

\item 
We have discussed the  spatial distribution of  stars in different age
intervals. We find  the distribution of young  stars to be clumpy  and
associated with the regions of high emission from the neutral gas, but
not coincident  with   the maxima  of   \hi\  column  density.   Older
main-sequence  stars  (with age    between    0.7  and  2  Gyr)    and
intermediate-age (RC and RGB) stars   show an extended, although   not
perfectly uniform,  distribution.  There  is  no obvious   correlation
between the distribution of these relatively old stars and that of the
neutral gas.

\item 
%- 
By  considering  the energy  input from  supernova explosions, we find
that the  ``hole'' in the \hi\ distribution  cannot have been produced
by a supernova driven wind  unless, 1) the  hole is unfeasibly old, 2)
supernovae are  able to  impart  2  orders  of magnitude  more kinetic
energy   to the ISM  than previously  thought, or  3) the  IMF is very
top-heavy. We can reasonably exclude 1 and 3, and 2 is very unlikely.
By far  the  most difficult  obstacle  for  the  supernova wind driven
hypothesis is the fact that the stellar  surface brightness within the
\hi\  hole is  lower than  that outside the   hole for any stellar age
range.   We find an appealing  alternative  to be that  the \hi\  ring
(rather than  ``hole'')  is the result  of  a recent accretion   of an
intra-group \hi\ cloud with a non-zero angular momentum.

\end{itemize}

%-----------------------------------------------------------------------
\begin{acknowledgements}

We  thank   the  anonymous  referee for   comments  that improved  the
presentation of  the paper.  We also  thank Dr.  L.   Young for kindly
providing us with her  \hi\ density maps,  Dr. L.  Girardi for  making
available  to us his set of  theoretical isochrones on the ACS system,
Dr.  E.   Brinks  for helpful  comments   on the  manuscript, and  Dr.
M. Sirianni for  communicating his results  on ACS characterization in
advance   of publication. EVH  wishes  to  thank the European Southern
Observatory for  hospitality during  a visit in  which this  paper was
partly written. YM acknowledges  financial  support by MIUR under  the
program PRIN2003.

\end{acknowledgements}

%------------------------------------------one column figure 
\begin{figure*}%[ht]
%\resizebox{\hsize}{!}{\includegraphics{./psfiles/F_izw18.ps}}
%\centering \includegraphics[width=15cm]{./psfiles/F_izw18.ps}
%\centering \includegraphics[width=15cm]{fig13.ps}
%\centering \includegraphics[width=18cm,height=18cm]{fig13.ps}
\centering \includegraphics[width=18cm]{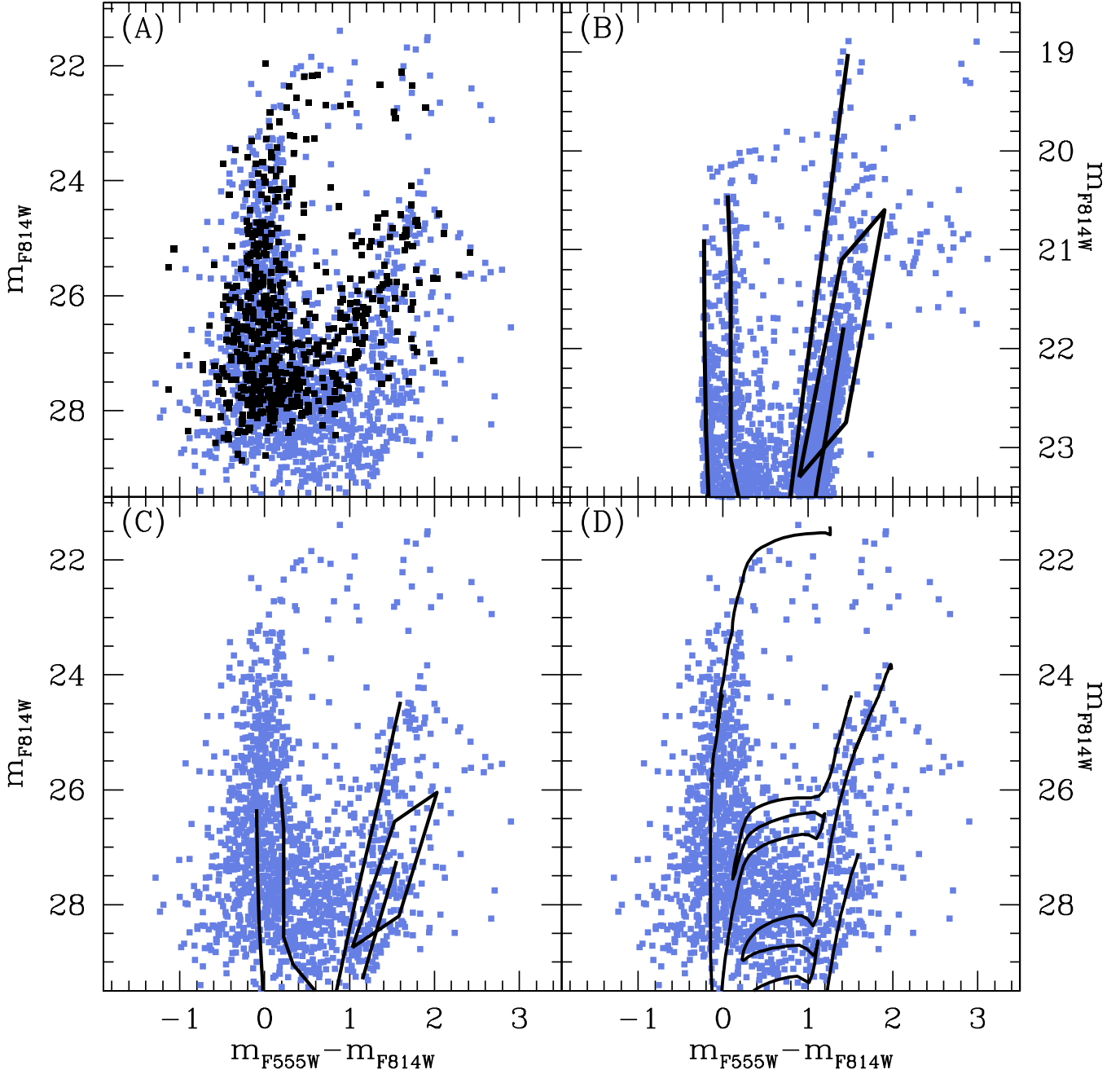}
\caption{The \ii, (\vviiZW) color-magnitude diagram of \zwi\ from ACS
archival data.  Panel~A compares the photometry of Izotov \& Thuan
(\cite{izo04}, {\it dark dots}) with the result of our reanalysis
({\it light dots}), showing a better recovery of the fainter red stars.
The HST/WFPC2 color-magnitude diagram of Sextans~A (Dolphin et al.
\cite{dolphin03}) is shown for comparison in Panel~B, together with
hand-drawn   fiducial lines marking  the  major evolutionary sequences
(MS, blue and red  super-giants, and RGB).   A box delimits red  giant
stars in the RGB tip region.
The same fiducial loci are superposed to  the CMD of \zwi\ in Panel~C,
corrected for the  distance and reddening  of Sextans~A and shifted to
the distance of \zwi, assuming \dmod~$  = 30.88$ (15 Mpc) with \ebv~$=
0.15$.   On the same assumptions,  theoretical isochrones from Girardi
et al. (2004, priv. comm.) with $Z=0.0004$ and ages 7, 80, 281 Myr and
10 Gyr are over-plotted in Panel~D.}
\label{f_izw18}
\end{figure*}
%-------------------------------------------------------------

%-----------------------------------------------------------------------

\listofobjects
%---------------------------------------------------------------------
%
\appendix

\section{On the age of \zwi}
\label{S_izw}

The discovery of a  population of presumably very   old stars (red  HB
stars) in SagDIG argues against  metal-poor galaxies being necessarily
{\em young}, i.e.  born in recent times.  In this  regard, we found it
of interest to re-examine the properties of the prototype of candidate
young galaxies, \zwi.

When  comparing the two  galaxies,  it should  be remembered  that the
metallicity difference between the two galaxies is not large: \zwi\ is
not so different after all from the  most metal-poor Local Group dwarf
galaxies    like   SagDIG   or   Leo~A  [see   Aloisi   et    al.
(\cite{aloisi03})     for a   recent compilation   of   heavy elements
abundances in \zwi].
%
% Infered abundances of  heavy elements, found in IZW18 (Aloisi 2003):
%
% [Fe/H] =  -1.76 ± 0.12
% [O/H]  =  -2.06 ± 0.28 
% [Si/H] =  -2.09 ± 0.12, [Ar/H] =  -2.27 ± 0.13, and [N/H] = -2.88± 0.11.
%
%
Yet \zwi\ is a key object in  that it lacks  a positive detection of a
red giant branch, which seems to imply that the galaxy is younger than
about 1 Gyr.  Even  the recent HST/ACS deep  study of Izotov  \& Thuan
(\cite{izo04}) was  unable to  trace  an ``old'' RGB.   These  authors
place \zwi\ further out than previously thought,  at 15~Mpc instead of
10~Mpc, and set an upper limit to the age of \zwi\ between 1$-$2 Gyr.

By     applying    the        reduction    methods     described    in
Sect.~\ref{S_observations}    to the  ACS   data   of Izotov  \& Thuan
(\cite{izo04}),  we have been  able to obtain a substantially improved
CMD   of \zwi.    The result   of   our reanalysis   is  presented  in
Fig.~\ref{f_izw18}, where  we show a   new \ii,  (\vviiZW) CMD  for  a
combined field containing  both the main body  and the ``C'' component
of \zwi.

%------------------OOO----------------------
%\clearpage

The new  diagram  is shown in  Panel~A,  along with that of  Izotov \&
Thuan (\cite{izo04}). The comparison  indicates an excellent match for
the    CMD features   in common,    with   no appreciable  photometric
differences between the two reductions. Our reduction however provides
access to a higher completeness everywhere, and deeper photometry.
This result  is probably to be attributed   to the updated geometrical
distortion correction and  pixel oversampling technique  employed here
(see   Anderson  \& King  \cite{jay00}) which  appears  to improve the
effectiveness of  PSF  fitting. Note that   the  selection criteria on
\allstar\ parameters are similar for both reductions.

%------------------OOO----------------------

It is for  the redder population  that the improvements in our diagram
are  most noticeable.  In   particular,  we can  identify    with some
confidence the  sequence of red  stars having colors (\vviiZW)~$=1-2$,
made of red super-giants.
This identification, crucial  to set observational constraints  on the
reconstruction of the SFH of \zwi, is supported by comparison with the
CMDs of  other dwarf irregular  galaxies.  Panel~B shows the HST/WFPC2
color-magnitude  diagram of  Sextans~A   obtained by  Dolphin  et  al.
(\cite{dolphin03}),  where we have marked  the main features (MS, blue
and red super-giants, and RGB) with  fiducial lines. The same fiducial
loci  are superimposed onto the  CMD  of \zwi\  in Panel~C, assuming a
distance of 15 Mpc for the galaxy.
While the higher luminosity reached by MS stars in \zwi\ is consistent
with its stronger star-burst, this comparison {\em shows an intriguing
agreement for the  older evolutionary sequences, in particular  around
the red super-giant and RGB tip regions}.
This conclusion is confirmed   by  a comparison with  isochrones  from
Girardi et al.  with $Z=0.0004$ and  ages 7, 80,  281 Myr, and 10 Gyr,
which are over-plotted on the data in Panel~D.

The termination point  of the  $10$ Gyr  isochrone  overlaps with  the
clump of stars seen just fainter than \ii~$\sim 27.3$.
Note also that the magnitude difference between the tip of the old RGB
and that of the RSG is typical of  dwarf irregulars, spanning 3.5--4.5
magnitudes in $I$ with some dependence on the SFH of each galaxy.
Although   this result does  not yet  represent a strong  case for the
identification  of an  ``old''  RGB, it seems  very encouraging.   Our
reanalysis gives supports to the 15 Mpc distance proposed by Izotov \&
Thuan (\cite{izo04}), and provides preliminary yet intriguing evidence
for a RGB  cutoff at \ii~$\sim 27.3$.  In  this scenario,  \zwi\ would
not appear substantially different from extremely metal-poor (yet old)
Local Group  dwarfs like  SagDIG  and Leo\,A.   Efforts to  push  even
deeper the  capabilities of ACS would  be  worthwhile to  confirm this
hypothesis.

%-------------------------------------------------------------------
\end{document}